\documentclass[reprint, amsmath, amssymb, aps, prx, twocolumn, linenumbers, notitlepage, superscriptaddress]{revtex4-2}
\usepackage{newtxtext,newtxmath}
\usepackage{graphicx}
\usepackage{dcolumn}
\usepackage{bm}
\usepackage[colorlinks=true,
 linkcolor=black,
 anchorcolor=black,
 citecolor=black,
 urlcolor=blue
 ]{hyperref}
\usepackage{verbatim}
\usepackage{braket}
\usepackage{siunitx}
{\rmfamily}
\usepackage{makecell}
\usepackage{xcolor}
\usepackage[switch]{lineno}

\makeatletter
\newcommand\sectionn[1]{%
  \vspace{1.5\baselineskip} 
  \noindent\textbf{\large #1}
  \par{}
  \vspace{0.5\baselineskip} 
}
\makeatother
\begin{document}
\nolinenumbers
\def\titl{Observation of non-Hermitian many-body phase transition in a Rydberg-atom array}
\title{\titl}
\author{Yao-Wen Zhang}
\thanks{These authors contributed equally to this work.}
\author{Biao Xu}
\thanks{These authors contributed equally to this work.}
\affiliation{
National Gravitation Laboratory, MOE Key Laboratory of Fundamental Physical Quantities Measurement,\\
Hubei Key Laboratory of Gravitation and Quantum Physics, PGMF,\\
Institute for Quantum Science and Engineering, School of Physics,\\
Huazhong University of Science and Technology, Wuhan 430074, China
}
\author{Yijia Zhou}
\thanks{These authors contributed equally to this work.}
\affiliation{Shanghai Qi Zhi Institute, Shanghai 200232, China}
\author{De-Sheng Xiang}
\thanks{These authors contributed equally to this work.}
\author{Hao-Xiang Liu}
\author{Peng Zhou}
\author{Kuan Zhang}
\author{Ren Liao}
\affiliation{
National Gravitation Laboratory, MOE Key Laboratory of Fundamental Physical Quantities Measurement,\\
Hubei Key Laboratory of Gravitation and Quantum Physics, PGMF,\\
Institute for Quantum Science and Engineering, School of Physics,\\
Huazhong University of Science and Technology, Wuhan 430074, China
}
\author{Thomas Pohl}
\email{thomas.pohl@itp.tuwien.ac.at}
\affiliation{Institute for Theoretical Physics, TU Wien, Vienna 1040, Austria}
\author{Weibin Li}
\email{weibin.li@nottingham.ac.uk}
\affiliation{School of Physics and Astronomy, and Centre for the Mathematics\\
and Theoretical Physics of Quantum Non-equilibrium Systems,\\
The University of Nottingham, Nottingham NG7 2RD, United Kingdom
}
\author{Lin Li}
\email{li\_lin@hust.edu.cn}
\affiliation{
National Gravitation Laboratory, MOE Key Laboratory of Fundamental Physical Quantities Measurement,\\
Hubei Key Laboratory of Gravitation and Quantum Physics, PGMF,\\
Institute for Quantum Science and Engineering, School of Physics,\\
Huazhong University of Science and Technology, Wuhan 430074, China
}
\affiliation{Wuhan Institute of Quantum Technology, Wuhan 430206, China}

\begin{abstract}
\bfseries
Non-Hermitian quantum mechanics with parity-time ($\mathcal{PT}$) symmetry offers a powerful framework for exploring the complex interplay of dissipation and coherent interactions in open quantum systems. While $\mathcal{PT}$-symmetry breaking has been studied in various physical systems, its observation on a quantum many-body level remains elusive. 
Here, we experimentally realize a non-Hermitian $XY$ model in a strongly-interacting Rydberg-atom array. By measuring the Loschmidt Echo of a fully polarized state, we observe distinct dynamical signatures of a $\mathcal{PT}$-symmetry-breaking phase transition. Dipole interactions are found to play a crucial role, not only determining the transition point but also triggering a non-Hermitian many-body blockade effect that protects the Loschmidt Echo from decay with a non-monotonic dependence on the system size. Our results reveal intricate interaction-induced effects on $\mathcal{PT}$-symmetry breaking and open the door for exploring non-Hermitian many-body dynamics beyond single-particle and mean-field paradigms. 
\end{abstract}

\maketitle

\begin{figure*}[!ht]
  \centering
  \includegraphics{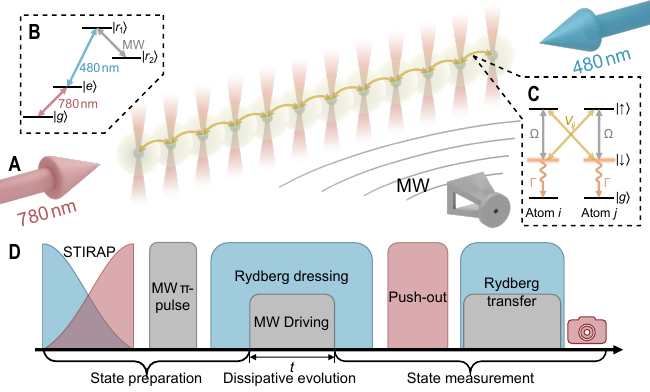}
  \caption{
  \textbf{Experimental protocol.} 
  (\textbf{A}) One-dimensional array of $^{87}$Rb atoms.  Atoms are initialized in the Rydberg state $\ket{\uparrow} = \ket{r_2} = \ket{69P_{3/2}, m_J=3/2}$. The state $\ket{\uparrow}$ is coupled to a dressed Rydberg state $\ket{\downarrow} = \tfrac{1}{\sqrt{2}}(\ket{e} + \ket{r_1})$ via a microwave (MW) field with Rabi frequency $\Omega$. The dressed state is a superposition of the Rydberg state $\ket{r_1} = \ket{68D_{5/2}, m_J=5/2}$ and a low-lying state $\ket{e}$. The insets show the electronic (\textbf{B}) and encoded (\textbf{C}) energy levels. By choosing state $\ket{e} = \ket{5P_{3/2}}$ or $\ket{6P_{3/2}}$, the effective decay rate of the dressed state $\ket{\downarrow}$ is $\Gamma = 2\uppi \times \SI{3.0}{MHz}$ or $2\uppi \times \SI{0.7}{MHz}$, respectively. The exchange interaction is controlled by varying the distance between the Rydberg atoms.  
  (\textbf{D}) Timing sequence. LE is measured via a two-step projective protocol: after the evolution, atoms in state $\ket{\downarrow}$ spontaneously decay to state $\ket{g}$ and are subsequently removed by a push-out laser, while those remaining in state $\ket{\uparrow}$ are transferred back to state $\ket{g}$ for fluorescence detection. 
  Successful detection of all atoms confirms that state $\ket{\uppsi_N(t)}$ returns to the initial state $\ket{0_N}$, giving the survival probability of the initial state, i.e., the LE, normalized with respect to the initial state (see Supplementary Materials).
  }
  \label{fig:fig1}
\end{figure*}

\noindent
Non-Hermitian physics, complementary to the conventional Hermitian quantum theory, provides a unique framework for understanding open quantum systems~\cite{ashidaNonHermitianPhysics2020}.
Although dissipative operators appear in their Hamiltonian, non-Hermitian systems exhibit real spectra and unitary dynamics when $\mathcal{PT}$ symmetry (PTS) is preserved~\cite{benderRealSpectraNonHermitian1998, el-ganainyNonHermitianPhysicsPT2018b}.
$\mathcal{PT}$-symmetry breaking (PTSB) occurs at exceptional points (EPs), signaled by complex-valued spectra~\cite{heissPhysicsExceptionalPoints2012, miriExceptionalPointsOptics2019a}. Broad explorations of these features have revealed rich physical phenomena~\cite{makrisBeamDynamicsMathcal2008,fengSinglemodeLaserParitytime2014a, hodaeiParitytimeSymmetricMicroring2014, xu2016topological,gongTopologicalPhasesNonHermitian2018, yaoNonHermitianChernBands2018} and suggest promising applications~\cite{assawaworrarit2017robust,wang2020petermann,kononchuk2022exceptional,selim2025selective}. 
Inspired by these prospects, non-Hermitian physics has been demonstrated in different settings~\cite{ruter2010observation,schindler2011experimental,regensburger2012parity,zhu2014pt,wu2019observation,naghiloo2019quantum,li2019observation,ding2021experimental,lin2022experimental,liang2023observation,zhaoTwodimensionalNonHermitianSkin2025}.
These seminal experiments have revealed exotic single-particle effects that are otherwise absent in Hermitian systems. 

On the other hand, recent theoretical work on the many-body regime predicts fascinating effects, such as Yang-Lee edge singularities~\cite{fisherYangLeeEdgeSingularity1978, shen_proposal_2023-1}, topological states~\cite{liuNonHermitianTopologicalMott2020,shenNonHermitianSkinClusters2022,yoshidaNonHermitianMottSkin2024,wanNonHermitianInteractingQuantum2025}, and phase transitions~\cite{matsumoto_continuous_2020, gopalakrishnan_entanglement_2021-1, lourencco2022non}, to emerge from the interplay of inter-particle interactions and dissipation. 
Experiments have discovered intriguing nonlinear behavior that emerges from mean field interactions in a non-Hermitian system of atomic quantum gases in optical microcavities~\cite{gaoObservationNonHermitianDegeneracies2015,ozturkObservationNonHermitianPhase2021}. Despite these advances, the realization and probing of $\mathcal{PT}$-symmetry breaking in a many-body system has remained an outstanding challenge.

In this work, we report such an experimental realization by implementing a non-Hermitian quantum spin chain in an array of strongly interacting Rydberg atoms. This is made possible by independently controlling dissipation and interaction, to engineer a fully tunable non-Hermitian $XY$ spin model. Through state-selective probing of the atoms, we measure the Loschmidt Echo (LE) of a polarized spin state as a hallmark for many-body $\mathcal{PT}$-symmetry-breaking transition. 
The LE exhibits a distinct time evolution, from highly oscillatory dynamics to exponential decay upon breaking $\mathcal{PT}$ symmetry. 
Remarkably, the LE displays an anomalous non-monotonic dependence on system size, which highlights the complex role of Rydberg interactions in the realized non-Hermitian spin chain. 
Moreover, we discover a non-Hermitian many-body blockade effect that protects quantum state against spreading and decay in a high-dimensional Hilbert space.

\sectionn{Non-Hermitian $XY$ model in a Rydberg atom array}
\noindent
In our experiment, individual $^{87}$Rb atoms are trapped in a one-dimensional programmable optical tweezer array (Fig.~\ref{fig:fig1}A). 
The atoms are laser-excited to Rydberg state $\ket{r_1}=\ket{68D_{5/2}, m_J=5/2}$ from the electronic ground state $\ket{g}$ via two-photon stimulated Raman adiabatic passage (STIRAP)~\cite{cubel2005coherent}. 
A microwave (MW) field transfers atoms from $\ket{r_1}$ to another Rydberg state $\ket{r_2}=\ket{69P_{3/2}, m_J=3/2}$ via a $\uppi$-pulse. 
State-dependent dissipation is engineered by coupling $\ket{r_1}$ to a low-lying state, $\ket{e}$, using a resonant laser. 
This creates a dressed state $\ket{\downarrow}=\tfrac{1}{\sqrt{2}}(\ket{e} + \ket{r_1})$ that decays to state $\ket{g}$ (Fig.~\ref{fig:fig1}, B and C). 
By selectively coupling the Rydberg state to either the state $\ket{e} = \ket{5P_{3/2}}$ or $\ket{6P_{3/2}}$, the decay rate of the dressed state can be tuned to $\Gamma = 2\uppi \times \SI{3.0}{MHz}$ or $2\uppi \times \SI{0.7}{MHz}$, respectively (see Supplementary Materials).
Spontaneous decay rates of the bare Rydberg states are much smaller than $\Gamma$ and thus have a negligible impact on the dynamics.
The states $\ket{\uparrow} = \ket{r_2}$ and $\ket{\downarrow}$ are then coupled using a resonant MW field with Rabi frequency $\Omega$.

Dynamics of states $\ket{\uparrow}$ and $\ket{\downarrow}$ are governed by a spin-1/2 many-body $XY$ Hamiltonian incorporating a local non-Hermitian operation, $H_\text{nh}=H_\text{pt}+H_\text{im}$. Here,
\begin{equation}
  H_{\text{pt}}= \sum_{i=1}^N \left( \frac{\Omega}{2} \upsigma^\text{x}_i + \text{i} \frac{\Gamma}{4} \upsigma^\text{z}_i \right) + \sum_{i<j}\frac{V_{ij}}{2} \left( \upsigma^\text{x}_i \upsigma^\text{x}_j + \upsigma^\text{y}_i \upsigma^\text{y}_j \right),
\end{equation}
where $\upsigma_i^{\text{x},\text{y},\text{z}}$ are Pauli operators of the $i$-th atom, correspondingly. The state-dependent dissipation generates an imaginary detuning $\text{i}\Gamma/2$ to each atom.
The dipolar exchange interaction between the Rydberg atoms, $V_{ij} = \tfrac{1}{2}C_3/r_{ij}^{3}$, depends on separation $r_{ij}$ between the $i$-th and $j$-th atoms, and resonant dipole coefficient $C_3$ (see Supplementary Materials). 
$H_{\text{im}}=-\text{i}\tfrac{N\Gamma}{4}\mathbb{I}$ is a homogeneous imaginary field, whose strength is linearly proportional to the system size $N$. The Hamiltonian $H_{\text{pt}}$ is $\mathcal{PT}$-symmetric, as $[\mathcal{PT}, H_{\text{pt}}] = 0$ (see Supplementary Materials). This can be verified using the parity operator $\mathcal{P} = \Pi_i \upsigma_i^x$, and the time-reversal operator $\mathcal{T}$, which performs complex conjugation~\cite{fringPTsymmetricDeformationsIntegrable2013}.

To explore the $\mathcal{PT}$-symmetric physics in the non-Hermitian dynamics, we employ the Loschmidt Echo---defined as the survival probability of the initial state $\ket{0_N}$ after time evolution~\cite{Gorin2006, heylDynamicalQuantumPhase2013, Yan2020, wong2022loschmidt}.
The experimental protocol is shown in Fig.~\ref{fig:fig1}D. The atoms are prepared in a fully polarized many-body state $\ket{0_N} = \ket{\uparrow\uparrow\cdots\uparrow}$. Subsequently, dissipative Rydberg dressing and MW driving are applied, and the system evolves under the non-Hermitian Hamiltonian $H_{\text{nh}}$.
After time $t$, LE is measured by projecting the time-evolved state $\ket{\uppsi_N(t)} = \exp(-\mathrm{i}H_{\text{nh}}t)\ket{0_N}$ back onto the initial state $\ket{0_N}$ and determining the survival probability $F(t) = \left| \braket{0_N | \uppsi_N(t)} \right|^2$ (see Supplementary Materials).
Experimentally, $F(t)$ corresponds to the probability of simultaneously detecting all atoms in $\ket{\uparrow}$, reflecting $N$-body coincidence events in the $2^N$-dimensional Hilbert space.

\begin{figure*}[!ht]
  \centering
  \includegraphics{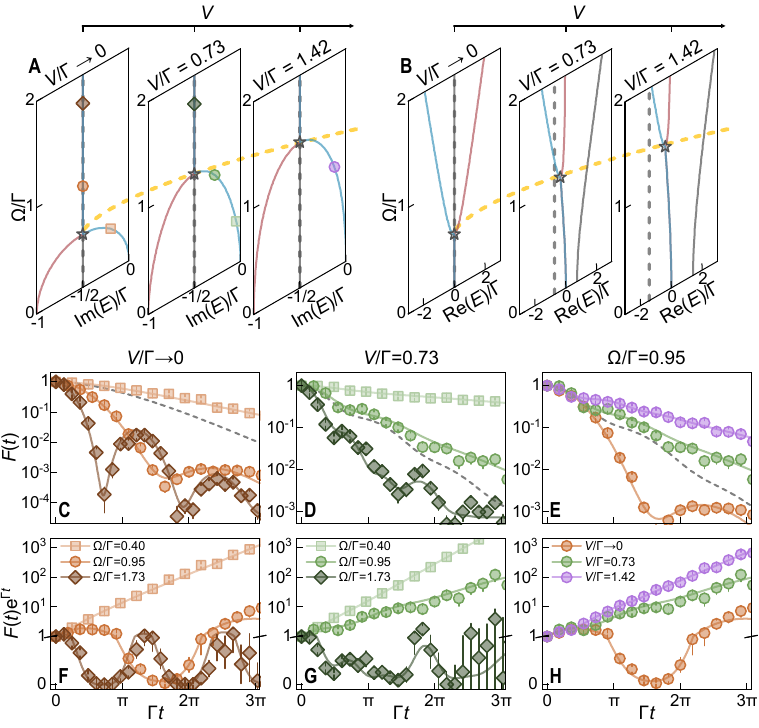}
  \caption{
  \textbf{Non-Hermitian spectra and dynamics of LE.} 
  Imaginary (\textbf{A}) and real (\textbf{B}) parts of the eigenvalues $E$. When $V = 0$, the four eigenstates are degenerate, while the non-negligible interaction $V$ lifts some of the degeneracy in Re$(E)$. The initial state $\ket{\uparrow\uparrow}$ does not couple to the anti-symmetric dark state $\propto (\ket{\uparrow\downarrow}-\ket{\downarrow\uparrow})$ (gray dashed line). 
  Im$(E)$ is shifted by $-\Gamma/2$ relative to the $H_\text{pt}$ spectrum due to $H_\text{im}$.The exceptional point (EP, star) shifts to higher $\Omega$ with increasing $V$, as given by the yellow dashed curve obtained from diagonalization of $H_\text{pt}$ (see Supplementary Materials).  
  (\textbf{C} to \textbf{E}) Evolution of the LE $F(t)$ for $V \to 0$ (C), $V = 0.73\Gamma$ (D), and $\Omega = 0.95\Gamma$ (E). The EPs are $\Omega_\text{c} = 0.50\Gamma$, $\Omega_\text{c} = 1.07\Gamma$, and $V_\text{c} = 0.50\Gamma$, respectively. Solid curves are numerical simulations using the experimental parameters, while dashed curves are simulated results at the EPs.  
  (\textbf{F} to \textbf{H}) Scaled LE, $F(t) \text{e}^{\Gamma t}$, corresponding to the data in (C to E). The scaled LE exhibits oscillatory (monotonic) behavior in the PTS (PTSB) phase. The experimental data agree well with numerical simulations (solid). To highlight the distinct dynamics in the PTS and PTSB regimes, linear scales are used for LE values between 0 and 1, and logarithmic scales for values above 1. Error bars represent the binomial standard deviation.}
  \label{fig:fig2}
\end{figure*}

\begin{figure*}[!ht]
  \centering
  \includegraphics{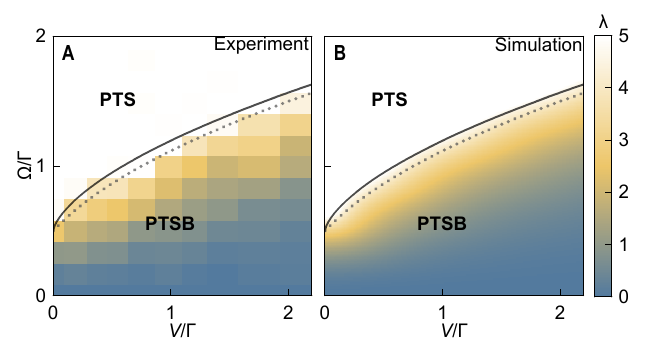}
  \caption{
  \textbf{$\mathcal{PT}$ phase diagram of interacting Rydberg atoms.} 
  Measured (\textbf{A}) and simulated (\textbf{B}) PTS--PTSB phase diagram. The color scale represents the rate function $\uplambda$ at the evolution time $\Gamma t_\text{e}=2.3\uppi$. The phase boundary is extracted at $\uplambda \approx 5$. The solid curve shows the phase boundary numerically obtained by analyzing spectra of the Hamiltonian $H_\text{pt}$, while the dotted line corresponds to the analytic approximation $\Omega_\text{c} \approx \Gamma \sqrt{1/4 + V/\Gamma}$.
  }
  \label{fig:fig3}
\end{figure*}

\sectionn{Exceptional point and Loschmidt Echo}
\noindent
To elucidate how the interactions induce striking non-Hermitian many-body effects, we first examine spectra and dynamics of two Rydberg atoms, shown in Fig.~\ref{fig:fig2}. The eigenvalues of $H_\text{nh}$ are numerically calculated and shown in Fig.~\ref{fig:fig2}, A and B. Through modulating the Rabi frequency and interaction strength $V$ between the atoms (via the atomic separation), the dynamics of the LE are measured, shown in Fig.~\ref{fig:fig2}, C to E. 
To access the dynamics governed by the $\mathcal{PT}$-symmetric Hamiltonian $H_{\text{pt}}$, the scaled LE, $F(t)\text{e}^{\Gamma t}$, is employed to compensate for the global decay induced by $H_{\text{im}}$, as presented in Fig.~\ref{fig:fig2}, F to H.

To realize the non-interacting regime, the atoms are placed approximately $\SI{114}{\um}$ apart, resulting in a negligible interaction strength of about $2 \uppi \times \SI{5}{kHz}$ compared to $\Gamma= 2 \uppi \times \SI{3.0}{MHz}$.
An exceptional point at $\Omega_\text{c}=\Gamma/2$ separates the PTS and PTSB phases.
In the PTS phase ($\Omega>\Omega_\text{c}$), the eigenvalues $E_{\pm}=\pm \sqrt{\Omega^2 - \Gamma^2/4} - \text{i}\Gamma/2$ of $H_{\text{nh}}$ are complex (Fig.~\ref{fig:fig2}, A and B): real parts have identical amplitude but opposite signs that cause oscillations of the LE, while imaginary parts induce damping at the constant rate $\Gamma/2$ (Fig.~\ref{fig:fig2}C, diamonds and circles). 
In the PTSB phase ($\Omega<\Omega_\text{c}$), $E_{\pm}$ are purely imaginary, such that LE decays with time (Fig.~\ref{fig:fig2}C, squares). 
Our initial state $\ket{\uparrow\uparrow}$ primarily projects onto the slower-decaying blue branch in Fig.~\ref{fig:fig2}A, resulting in the decay rate lower than that in the PTS phase.

To investigate the interacting case, the distance between the atoms is reduced to \SI{15}{\um}, yielding a dipolar exchange interaction strength $V=2\uppi\times\SI{2.2}{MHz}=0.73\Gamma$. 
The interaction shifts the EP to a higher value, as depicted in Fig.~\ref{fig:fig2}, A and B.
An analytical EP can be approximately obtained as: $\Omega_\text{c}\approx \Gamma\sqrt{\tfrac{1}{4}+\tfrac{V}{\Gamma}}$ (see Supplementary Materials). The approximated $\Omega_\text{c}$ agrees well with the exact numerical result for a wide range of parameters, which will be discussed later in Fig.~\ref{fig:fig3}. 

The interaction lifts the eigenmode degeneracy, resulting in three distinct branches relevant to our exchange-symmetric initial state $\ket{\uparrow\uparrow}$ (Fig.~\ref{fig:fig2}, A and B), leading to rich dynamical features.
In the PTS region, the three eigenvalues have distinct real parts, contributing to more complex, non-single-frequency oscillations. This is evident in the LE dynamics, which exhibits irregular oscillatory patterns (Fig.~\ref{fig:fig2}D, diamonds). These complicated patterns arise from the interference of different frequencies when the respective modes are excited, as the interaction $V$ couples the two-atom basis states.
In the PTSB region, the interaction decreases the magnitude of the imaginary part of the blue branch in Fig.~\ref{fig:fig2}A, leading to slower decay (Fig.~\ref{fig:fig2}D, squares and circles) compared to the non-interacting case. Additionally, the Rydberg atom interactions shift the real parts of eigenmodes away from zero, causing weak oscillations in contrast to the previously pure decay dynamics.

\sectionn{$\mathcal{PT}$-symmetry breaking by dipole-dipole interactions}
\noindent
The dipolar interaction between the Rydberg atoms emerges as a pivotal parameter for driving the dynamical phases.
When $\Omega>{\Gamma}/{2}$, there is a critical interaction strength $V_{\text{c}}$ that marks the boundary between the PTS and PTSB phases.
To investigate this interaction-driven transition, experimental measurements of the LE are performed by varying $V$ with a fixed $\Omega = 0.95\Gamma$.
As shown in Fig.~\ref{fig:fig2}E, the LE changes from damped oscillations (orange circles) in the PTS phase into monotonic decay (green and purple circles) in the PTSB phase, as $V$ changes across the critical point $V_{\text{c}} = 0.50\Gamma$ (see Supplementary Materials). 

Here, a key distinction between the two phases is that the LE is much smaller in the PTS than in the PTSB regime. This qualitative difference in dynamical patterns allows us to probe the dynamical phase diagram arising from the interplay among interaction, driving, and dissipation. 
To this end, the Loschmidt rate function, $\uplambda(t_\text{e})=-\ln F(t_\text{e})$, is obtained from the experimental data for a range of interaction $V$ and Rabi frequency $\Omega$, shown in Fig.~\ref{fig:fig3}A. Here $t_\text{e}$ is chosen sufficiently long ($\Gamma t_\text{e}=2.3\uppi\gg1$) such that the LE dynamics are dominated by exponential decay (Fig.~\ref{fig:fig2}, C to E)), leading to a significantly larger decay rate $\uplambda(t_\text{e})$ in the PTS phase. From the perspective of $H_\text{pt}$, this longtime evolution ensures that the dynamics are governed by the exponential gain in the PTSB phase and the oscillation in the PTS phase, as shown by the scaled LE (Fig.~\ref{fig:fig2}, F to H).

\begin{figure*}[!ht]
  \centering
  \includegraphics{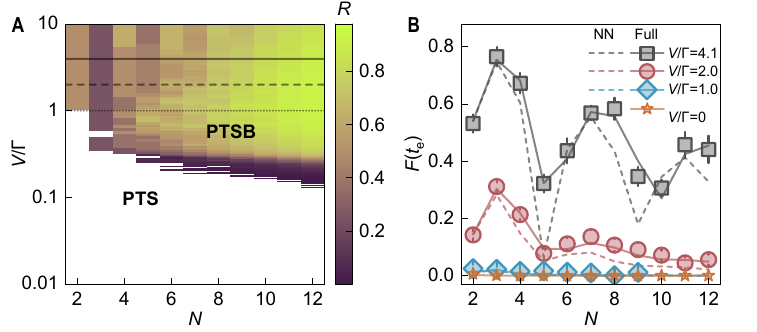}
  \caption{
  \textbf{Non-Hermitian phase transition and LE of $N$ Rydberg atoms.} 
  (\textbf{A}) Many-body PTS--PTSB phase diagram. In PTSB phases, the participation ratio $R$, i.e., the fraction of eigenvalues that are imaginary in the spectra, is nonzero. The ratio varies with the system size $N$. 
  (\textbf{B}) The measured LE value, $F(t_{\text{e}})$, as a function of $N$.  The solid and dashed lines show numerical simulations with full-range and nearest-neighbor (NN) interactions, respectively. The full-range model agrees well with the experimental data for all $N$ and interaction strengths, underscoring the essential role of long-range interactions. For the interacting cases ($V/\Gamma = 4.1, 2.0, 1.0$), the corresponding positions in the phase diagram (A) are marked by solid, dashed, and dotted lines, respectively. Data for $V/\Gamma=1.0$ are measured for up to 9 atoms, limited by the field of view of the atom trapping and imaging system. Data points for $V/\Gamma = 0$ are inferred from single-atom measurements. Parameters used in the experiment are: $\Gamma = 2\uppi \times \SI{0.7}{MHz}$, $\Omega/\Gamma = 1.2$, and $t_\text{e} = \SI{1.5}{\us}$. Error bars indicate the binomial standard deviation.
  }
  \label{fig:fig4}
\end{figure*}

In Fig.~\ref{fig:fig3}A, we observe a clear PTS--PTSB phase boundary with a threshold value $\uplambda(t_\text{e})\approx 5$. The measured phase diagram reveals an interaction-induced shift of the phase transition point. In stark contrast to the non-interacting systems where the phase boundary is fixed, here the critical driving strength for $\mathcal{PT}$-symmetry breaking is observed to increase with the interaction strength, highlighting a unique many-body effect on the system's non-Hermitian properties.
For a given $V$, the decay rate increases with $\Omega$, consistent with the growing magnitude of $|\text{Im}(E)|$ in the blue branch of the PTSB phase (Fig.~\ref{fig:fig2}A). 
Our data confirm that the decay rate in the PTSB phase is determined by the corresponding eigenmode, whereas it remains $\Gamma/2$ in the PTS phase, aligning with the spectral analysis. The numerical simulations based on the non-Hermitian Hamiltonian $H_\text{nh}$ and experimental parameters (see Supplementary Materials) match well with the measured phase diagram, as shown in Fig.~\ref{fig:fig3}B. This excellent agreement reflects the high-precision control over the atom array, MW and laser fields, as well as the dipolar interaction in our experiment.

\sectionn{Non-Hermitian phase transition in many-body system}
\noindent
Increasing the number of Rydberg atoms $N$ qualitatively reshapes the phase structure of the interacting non-Hermitian system. In particular, the PTS--PTSB phase boundary depends sensitively on $N$ (see Supplementary Materials), reflecting genuine many-body effects.
To explore the multi-atom $\mathcal{PT}$ phase transition, we first diagonalize the Hamiltonian $H_{\text{pt}}$ under a typical driving strength $\Omega/\Gamma=1.2$ with different $V/\Gamma$ and $N$.
The phase boundary is determined by analyzing the participation ratio of the symmetry-breaking modes---i.e., eigenstates with complex eigenenergies~\cite{hamazakiNonHermitianManyBodyLocalization2019}---defined as $R=D_{\text{I}}/D$, where $D_{\text{I}}$ is the number of symmetry-breaking modes and $D=2^N$ is the Hilbert-space dimension. 
The theoretical phase diagram is shown in Fig.~\ref{fig:fig4}A, with the PTS region ($R=0$) marked in white.
The critical interaction strength governing the $\mathcal{PT}$-symmetry-breaking transition, $V_{\mathrm c}$, falls steeply with $N$, from $V_{\mathrm c}\approx \Gamma$ for $N=2$, to $V_{\mathrm c}\approx 0.1\Gamma$ for $N>10$ (Fig.~\ref{fig:fig4}A), as numbers of collective symmetry-breaking modes increase significantly in larger systems. In the PTSB phase, although $R$ exhibits an overall increasing trend with $N$ at fixed $V$, its behavior is non-monotonic and shows pronounced oscillations with $N$, especially in the strongly-interacting regime ($V/\Gamma > 1$). 
This oscillatory size dependence underscores the role of interactions in the emergent multi-atom phase diagram.

To experimentally probe the size-dependent effects in the PTSB phase in Fig.~\ref{fig:fig4}A, a linear array of $N$ Rydberg atoms is prepared. In this multi-atom configuration, the Rydberg state $\ket{r_1}$ is coupled to state $\ket{e} = \ket{6P_{3/2}}$, reducing the decay rate of the dressed state $\ket{\downarrow}$ to $\Gamma = 2\uppi \times \SI{0.7}{MHz}$. This enables increasing $V/\Gamma$ without placing atoms too close, and hence preventing excessive fluctuations in $V$.
Starting from the initial state $\ket{0_N} = \ket{\uparrow\uparrow\cdots\uparrow}$, we let the system evolve over a sufficiently long duration $t_{\text{e}}$, such that $\Gamma t_{\text{e}} = 2.1\uppi \gg 1$, and then measure the LE.
 
Figure~\ref{fig:fig4}B shows the measured LE as a function of $N$ for different interaction strengths.
In the PTS phase ($V \ll \Gamma$), $F(t_{\text{e}})$ remains small---consistent with the linear increase in decay rates with $N$ due to the homogeneous imaginary field $H_{\text{im}} \propto N$, which reduces amplitudes of the LE at long times.
In contrast, in the PTSB phase, the LE not only reaches substantially higher values but also develops a pronounced non-monotonic dependence on $N$, as detailed below.

\begin{figure*}[!ht]
  \centering
  \includegraphics{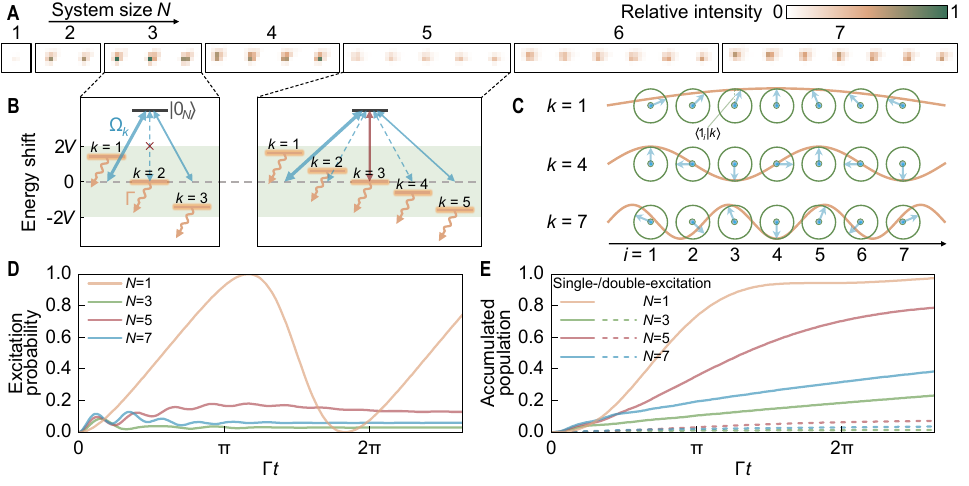}  
  \caption{
  \textbf{Non-Hermitian blockade in the $N$-atom chain.} 
  (\textbf{A}) Synthesized images of the $N$-atom chain. The color scale indicates the relative intensity of atomic fluorescence, rescaled by the corresponding measured LE in Fig.~\ref{fig:fig4}B.
  (\textbf{B}) Energy $U_k$ (orange) and coupling strength $\Omega_k$ of the spin-wave states. Arrow thickness shows the coupling strength $\Omega_k$, with dashed arrows indicating $\Omega_k = 0$. The red arrow marks the resonant case $U_k = 0$.  
  (\textbf{C}) Illustration of the spin-wave function (orange) for $N=7$. The site-dependent coupling $\braket{1_i|k}$ is represented by the vertical projection of the blue arrows. The wave function oscillates with increasing $k$. 
  (\textbf{D}) Non-Hermitian blockade dynamics. Simulated excitation probability versus evolution time $t$ for different system sizes (see Supplementary Materials). The excitation probability is calculated by $1-\tilde{F}(t)$, where $\tilde{F}(t)$ denotes the LE normalized by the total population remaining in the Hilbert space $\left\{\ket{\uparrow},\ket{\downarrow}\right\}^{\otimes N}$ at $t$. Excitation is strongly suppressed for $N>1$ cases. Parameters: $V/\Gamma = 4.1$; other parameters as in Fig.~\ref{fig:fig4}B.
  (\textbf{E}) Suppression of multi-excitations. Accumulated population loss, defined as $\Gamma\int_0^t p_m(t')  {\rm d}t'$, for the single- ($m=1$; solid lines) and double-excitation ($m=2$; dashed lines) manifolds. Here, $p_m(t')$ represents the instantaneous population in the $m$-excitation manifold at time $t'$. The significantly lower populations in the double-excitation manifold show the quantum Zeno-like mechanism induced by dissipation.
  }
  \label{fig:fig5}
\end{figure*}

First, for sufficiently strong interactions ($V > \Gamma$), the measured LE is significantly enhanced compared to the non-interacting case. For instance, at $V=2\Gamma$, $F(t_{\text{e}})$ becomes clearly visible, and at $V=4.1\Gamma$ it rises to much higher values. In particular, even at $N=12$, the LE remains substantial, $F(t_{\text{e}})=$ 0.44(5), which is highly non-trivial. For an open quantum system with a Hilbert space of dimension $2^{12}$, one would typically expect the survival probability to be negligible after prolonged evolution due to state spreading and dissipation. The anomalously high LE observed here suggests that the interplay of dissipation and interactions in the many-body PTSB phase strongly protects the initial quantum state against delocalization and decay.

Moreover, beyond the overall enhancement, the LE exhibits striking oscillatory patterns with $N$. Local maxima occur at $N = 3$ and 7, and minima at $N = 5$ and 10, forming a clear non-monotonic profile that strengthens with $V$ (e.g., $V=4.1\Gamma$). This oscillatory dependence is rather counterintuitive: generally, the survival probability of a many-body state is expected to decay rapidly---typically exponentially---with increasing system size, instead of exhibiting the oscillations. The observed anomalous size dependence suggests the emergence of a previously unexplored many-body dynamical mechanism. Its origin is not immediately apparent from the phase diagram in Fig.~\ref{fig:fig4}A alone, which is purely based on the eigenspectra. This motivates us to develop the spin-wave analysis in the following section that provides a microscopic understanding of these novel dynamics.

\sectionn{Non-Hermitian blockade of spin-wave states}
\noindent
We attribute the observed high LE values and their dependence on $N$ to an emergent \textit{non-Hermitian many-body blockade} effect. The microscopic mechanism behind this phenomenon can be broken down into the level structure that closely depends on the system size $N$ (Fig.~\ref{fig:fig5}, A and B), where the Rydberg atoms form collective spin-wave states due to the long-range exchange interaction (see Supplementary Materials). In terms of the spin-wave basis of the $XY$ model, state $\ket{0_N}$  is the vacuum state of the spin $\ket{\downarrow}$. It dominantly couples to spin-wave states consisting of a single excitation, $\ket{k} = \sqrt{\tfrac{2}{N+1}} \sum_{i=1}^N \sin\left(\tfrac{ik\uppi}{N+1}\right) \ket{1_i}$ with state $\ket{1_i}=\upsigma^\text{x}_i\ket{0_N}$. The coupling dynamics are described by an effective Hamiltonian,
$H_\text{eff} \approx \sum_k \left[-( U_k + \text{i}\tfrac{\Gamma}{2} )\ket{k}\bra{k} + \tfrac{\Omega_k}{2} ( \ket{0_N}\bra{k} + \ket{k}\bra{0_N} )\right]$, where $U_k$ and $\Omega_k$ are spin-wave energy
and coupling strength, respectively (Fig.~\ref{fig:fig5}B) (see Supplementary Materials). Moreover, all eigenenergies have an identical imaginary part $\text{i}\Gamma/2$, which causes decay of the excited spin-wave states. 

To understand the non-Hermitian blockade and the observed $N$-dependence of the LE in Fig.~\ref{fig:fig4}B, we analyze the energy and coupling strength of the spin-wave states.
Spin-wave energy $U_k = 2V\cos\left(\tfrac{k\uppi}{N+1}\right)$ depends on whether $N$ is even or odd non-trivially. 
For even $N$, all states $\ket{k}$ are shifted off resonance when $V \gg \Omega$, suppressing spin-wave excitation and leading to the higher LE seen in Fig.~\ref{fig:fig4}B.
For odd $N$, the state $\ket{k = (N+1)/2}$ is energetically resonant with state $\ket{0_N}$, but its coupling $\Omega_k = \Omega[1-(-1)^k] \sqrt{\tfrac{2}{N+1}} \cot\left(\tfrac{k\uppi}{2N+2}\right)$ varies strongly with $N$. When $N = 3, 7, 11, \dots$ (with a period of 4), the state is spatially anti-symmetric, which diminishes the coupling.
Figure~\ref{fig:fig5}C illustrates the spatial profile of the $k=4$, $N=7$ spin-wave state.
The vanishing coupling results in the high LE values and local maxima observed at these $N$ under $V = 4.1\Gamma$ in Fig.~\ref{fig:fig4}B.
In contrast, for $N = 5, 9, 13, \dots$, the coupling becomes non-negligible, allowing spin-wave excitation and reducing the LE---consistent with the measured local minima in the experiment.

The non-Hermitian blockade is further illustrated by the simulated excitation probability for different system sizes (Fig.~\ref{fig:fig5}D). 
Compared to the single-particle case, the excitation in an interacting Rydberg-atom chain is significantly suppressed, analogous to the conventional Rydberg blockade. 
However, it exhibits strong $N$-dependence, highlighting the many-body nature of the non-Hermitian blockade effect.

We emphasize that the observed non-Hermitian blockade originates from the long-range dipolar interaction. Simulations using the full interaction quantitatively match the experimental data in Fig.~\ref{fig:fig4}B, whereas a nearest-neighbor (NN) approximation only reproduces the qualitative trend. We also note that while the simulations in Figs.~\ref{fig:fig2}--\ref{fig:fig5} use a non-Hermitian Schrödinger equation, the results are equivalent to the master-equation approach, since the LE is a measure of the probability in the no-jump subspace (see Supplementary Materials).

The suppression of the multiple excitation may relate to the quantum Zeno-like effect~\cite{itano1990quantum,chen2025collective}, where dissipation acts as a continuous measurement of the initial state and singly-excited spin-wave states (see Supplementary Materials).
The simulation in Fig.~\ref{fig:fig5}E shows that, double excitations are strongly suppressed, since frequent measurements (large $\Gamma$) effectively confine the dynamics to the initial and single-excitation subspaces (see Supplementary Materials). The non-Hermitian many-body blockade thus emerges from the interplay of this Zeno-like effect and dipolar exchange interactions, further distinguishing it from conventional Rydberg blockade.

\sectionn{Summary and outlook}
\noindent
In summary, we have realized quantum simulation of non-Hermitian many-body physics by programming the geometry, dipolar interaction, and dissipation in a Rydberg atom array.
We have observed rich non-Hermitian phenomena in the quantum many-body regime, ranging from interaction-driven $\mathcal{PT}$-symmetry-breaking transition with anomalous system-size dependence, to the robust protection of quantum state against spreading and decay in large Hilbert space, to the emergent non-Hermitian many-body blockade effect. These results establish a dynamic pathway to probe non-Hermitian physics beyond the single-particle and mean-field paradigms.

Our $XY$ spin model can be readily extended to the $XXZ$ type with advanced Hamiltonian engineering techniques~\cite{scholl2022microwave}, paving the way for observing complex Berry phases~\cite{tsubotaSymmetryprotectedQuantizationComplex2022} and topological excitations~\cite{chenTopologicalSpinExcitations2023} in a non-Hermitian Hamiltonian.
Looking further ahead, as Rydberg-atom quantum simulators advance---for instance, through the deep integration of analog evolution and digital circuits---this progress will enable the exploration of a broader range of exotic non-Hermitian physics.
Specifically, phenomena such as topological states~\cite{liuNonHermitianTopologicalMott2020,shenNonHermitianSkinClusters2022,yoshidaNonHermitianMottSkin2024,wanNonHermitianInteractingQuantum2025}, many-body quantum chaos~\cite{zhouDiagnosingQuantumManybody2025}, Dirac fermions~\cite{yuNonHermitianStronglyInteracting2024}, and quantum scars~\cite{shenEnhancedManyBodyQuantum2024a} in many-body non-Hermitian systems may come within experimental reach.

\vspace*{0.5\baselineskip}

\bibliographystyle{sciencemag}
\bibliography{ref}

\newpage

\noindent
\textbf{\large Data availability}

\noindent
Data that support the figures within this paper are available from the corresponding author upon reasonable request.

\vspace*{1\baselineskip}

\noindent
\textbf{\large Code availability}

\noindent
The code used in this study is available from the corresponding author upon reasonable request. 

\vspace*{1\baselineskip}

\noindent
\textbf{\large Acknowledgments}

\noindent
The authors acknowledge insightful discussions with David Petrosyan, Igor Lesanovsky, Li You, Le-Man Kuang, and Tao Shi.
This work was supported by the National Key Research and Development Program of China (Grant No.~2021YFA1402003), the National Science and Technology Major Project of the Ministry of Science and Technology of China (Grant No.~2023ZD0300901), the National Natural Science Foundation of China (Grant Nos.~12374329, U21A6006 and 12404580), the Science and Technology Commission of Shanghai Municipality (Grant No. 25LZ2601002). Y.Z. acknowledges support from the Shanghai Qi Zhi Institute Innovation Program SQZ202317. T.P. acknowledges supported from the Austrian Science Fund (Grant No. 10.55776/COE1) and the European Union (NextGenerationEU). W.L. acknowledges support from the EPSRC through Grant No. EP/W015641/1. 

\vspace*{1\baselineskip}

\noindent
\textbf{\large Author contributions}

\noindent
All authors contributed substantially to this work.

\vspace*{1\baselineskip}

\noindent
\textbf{\large Competing interests}

\noindent
The authors declare no competing interests.

\clearpage
\onecolumngrid
\makeatletter
\renewcommand{\fnum@figure}{\small\textbf{fig. \thefigure}}
\renewcommand{\fnum@table}{\small\textbf{table \thetable}}
\makeatother
\renewcommand\thesection{S\arabic{section}}
\renewcommand\thesubsection{S\arabic{section}.\arabic{subsection}}
\renewcommand\thesubsubsection{S\arabic{section}.\arabic{subsection}.\arabic{subsubsection}}
\renewcommand{\thefigure}{S\arabic{figure}}
\renewcommand{\thetable}{S\arabic{table}}
\renewcommand{\theequation}{S\arabic{equation}}
\renewcommand{\thepage}{S\arabic{page}}
\setcounter{figure}{0}
\setcounter{table}{0}
\setcounter{equation}{0}
\setcounter{page}{1}
\begin{minipage}{0.95\textwidth}
\begin{center}
\Large\bfseries Supplementary Materials for ``\titl''
\end{center}
\tableofcontents
\end{minipage}

\section{Experimental methods}
\noindent
Non-Hermitian physics has been extensively studied across a wide range of experimental platforms, such as photonic systems, electronics, acoustics, superconducting circuits, cold and thermal atomic ensembles, NV centers, NMR, and trapped ions~\cite{ruter2010observation,schindler2011experimental,regensburger2012parity,peng2014parity,zhu2014pt,zhou2018observation,wu2019observation,naghiloo2019quantum,li2019observation,yu2020experimental,xiao2020non,ding2021experimental,wang2021topological,chen2021quantum,chen2022decoherence,lin2022experimental,lin2022non,liang2023observation,cao2023probing,wu2024third,daiNonHermitianTopologicalPhase2024,wang2025non,zhaoTwodimensionalNonHermitianSkin2025,lu2025dynamical,zhang2025observation,shen2025observation,salcedo2025demonstration}.
Programmable atom array~\cite{piotrowicz2013two,kim2016situ,endres2016atom,barredo2016atom,manetsch2024tweezer,lin2025ai} has emerged as a transformative platform for quantum simulation, enabling the study of strongly correlated phases and quantum dynamics in a wide range of models~\cite{bernien2017probing,kim2018detailed,de2019observation,shaw2024benchmarking,fang2024probing,xiang2024observation,chen2025interaction,zhao2025observation,liang2025observation,zhang2025observationwenlanchen}. 
Yet, the exploration of many-body non-Hermitian physics with Rydberg-atom arrays remains an open frontier.
In this work, we present an experimental realization of a non-Hermitian $XY$ spin model using a Rydberg-atom array. This section describes the preparation of the atom array, timing sequences, state preparation and measurement procedures, as well as the tuning and calibration of parameters and efforts to minimize evolution errors.

\subsection{Preparation of the atom array}
\noindent
The experiment is performed on $^{87}$Rb atoms trapped in optical tweezers. An optical tweezer array is generated using a spatial light modulator (SLM, HED 6010-NIR-080-C, Holoeye) and homogenized via feedback based on real-time imaging. The tweezers are loaded from a magneto-optical trap (MOT) followed by a rearrangement stage, where the target sites are fully loaded using the atoms trapped in the reservoir sites with a moving tweezer generated by a pair of acousto-optic deflectors (AODs, DTSX-400-800.850, AA Opto-Electronic) driven by an arbitrary waveform generator (AWG, M4i.6631-x8, Spectrum) with two output channels according to the results of fluorescence imaging before the rearrangement. After the rearrangement, a second fluorescence imaging is performed to ensure a defect-free array, i.e., all target sites occupied and no unwanted atoms remaining. The pattern of the tweezers and the target array are fully programmable, which enables tuning of the dipole-dipole interactions between Rydberg atoms by changing their separations. In this work, a one-dimensional chain of $N$ atoms is set as the target array, with all atoms equally spaced at a distance $r$. The total length of the atomic chain $(N-1)r$ is limited by the field of view of the microscopic imaging system.

\begin{figure}
  \centering
  \includegraphics[width=\textwidth]{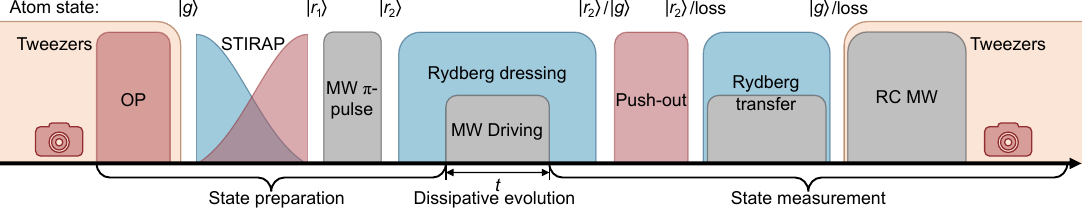}
  \caption{
    \textbf{Experimental sequence.} 
    After arranging the atom array, we perform optical pumping (OP) using a $\upsigma^{+}$-polarized \SI{795}{nm} laser that prepares the atoms in the ground state $\ket{5S_{1/2},F=2,m_F=2}$ (denoted $\ket{g}$). The atoms are then transferred to state $\ket{68D_{5/2},m_J=5/2}$ ($\ket{r_1}$) via the two-photon STIRAP scheme, followed by a resonant microwave (MW) pulse that transfers the atoms to state $\ket{69P_{3/2},m_J=3/2}$ ($\ket{r_2}$). A resonant \SI{480}{nm} Rydberg-dressing laser that couples states $\ket{r_1}$ to $\ket{5P_{3/2},F=3,m_F=3}$ ($\ket{e}$) clears residual population in state $\ket{r_1}$, ensuring that the non-Hermitian dynamics start with all atoms in state $\ket{r_2}$.  
    Subsequently, the resonant \SI{480}{nm} laser couples state $\ket{r_1}$ to $\tfrac{1}{\sqrt{2}}(\ket{r_1}\pm\ket{e})$ with Rabi frequency $\Omega_{480}$, while the MW field resonantly couples state $\ket{r_2}$ with $\tfrac{1}{\sqrt{2}}(\ket{r_1}+\ket{e})$ with strength $\Omega$. After an evolution time $t$ under these two fields, a projective measurement reads out population in state $\ket{r_2}$. The sequence consists of a resonant \SI{480}{nm} read-out pulse, a resonant \SI{780}{nm} push-out pulse that removes atoms in state $\ket{g}$, and a final step where the \SI{480}{nm} and MW fields are applied simultaneously to transfer state $\ket{r_2}$ to $\ket{g}$. Residual Rydberg atoms are cleared by a frequency-multiplied broadband Rydberg-clearing (RC) MW pulse with multiple sidebands.  
    During all stages involving Rydberg states, the optical tweezers are turned off. Final fluorescence imaging reveals the measurement outcome: bright sites correspond to atoms in state $\ket{r_2}$ at the end of the evolution, while dark sites indicate loss.
  }
  \label{fig:sequence}
\end{figure}

\subsection{State preparation and measurement}
\noindent
A typical experimental sequence is shown in fig.~\ref*{fig:sequence}. For the state preparation, non-Hermitian evolution and the state measurement stages, the quantization axis is defined by a magnetic field of \SI{30}{G} applied along the atom chain. The magnetic field is switched off during the fluorescence imaging stages that occur before state preparation and after the non-Hermitian evolution to avoid shifting the atomic transition frequencies and ensure clear imaging.

\begin{figure}
  \centering
  \includegraphics{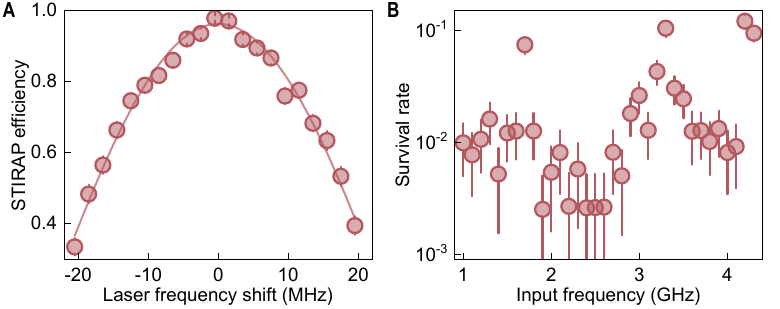}
  \caption{
    \textbf{STIRAP and Rydberg-clearing MW spectra.} 
    (\textbf{A}) Measured survival probability after the STIRAP, push-out, and read-out sequences as a function of the \SI{480}{nm} laser frequency. The fitted Gaussian curve yields a peak transfer efficiency of $\sim$95\%.  
    (\textbf{B}) Measured survival probability (dark counts) after the STIRAP and Rydberg-clearing sequences as a function of the MW frequency at the input of the frequency multiplier. The dark-count probability at the experimental condition of a \SI{2.5}{GHz} input frequency is 0.003(2). Error bars denote the binomial standard deviation.
  }
  \label{fig:STIRAP&clear}
\end{figure}

All atoms are prepared in state $\ket{r_2}$ using a two-stage scheme. First, the STIRAP (stimulated Raman adiabatic passage) scheme~\cite{cubel2005coherent} excites the atoms from state $\ket{g}$ to $\ket{r_1}$ with an efficiency of around 95\% (fig.~\ref*{fig:STIRAP&clear}A). Then the MW pulse accomplishes the transfer from state $\ket{r_1}$ to $\ket{r_2}$ with a probability that depends on the interaction strength $V$. This leads to a $V$-dependent state preparation efficiency that significantly affects the measurement of Loschmidt Echo (LE), denoted as $F$. The measured LE is therefore normalized by $F(V,\Omega,t)=\tfrac{F_\text{raw}(V,\Omega,t)}{F_\text{raw}(V,\Omega,t=0)}$ or $F(V,\Omega,t)=\tfrac{F_\text{raw}(V,\Omega,t)}{F_\text{raw}(V,\Omega=0,t)}$ with raw data $F_\text{raw}$. 

After the Rydberg state transfer process, a Rydberg-clearing (RC) MW pulse is applied to clear out all atoms remaining in Rydberg states that contribute to dark counts. It has a high clearing efficiency of $\sim99.7\%$ (fig.~\ref*{fig:STIRAP&clear}B). The overall dark counting is around 0.005, arising from the Rydberg-state atoms which survive the RC MW pulse and decay into the ground state, and the ground-state atoms which survive the push-out process. 
This dark-count effect is taken into account in numerical simulations and fits well in Fig.~\ref{fig:fig2}C.

\begin{figure}
  \centering
  \includegraphics{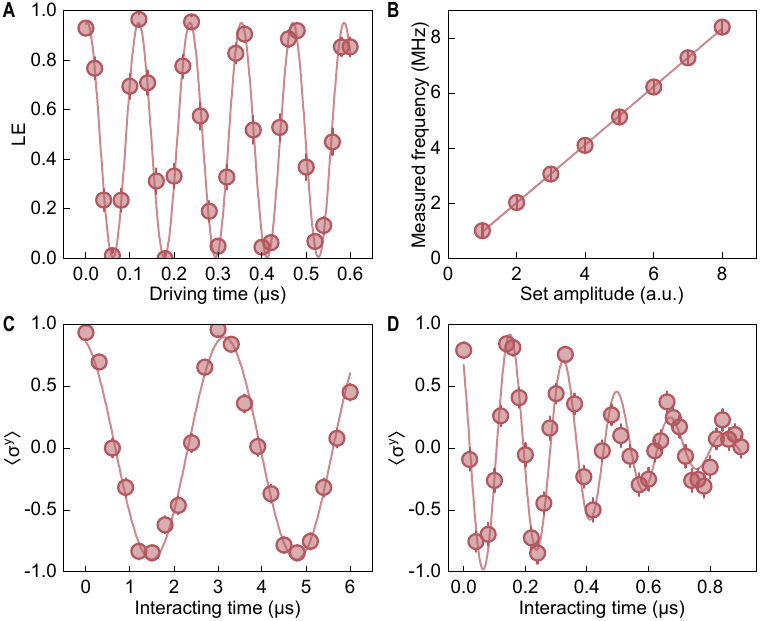}
  \caption{
    \textbf{Parameter calibration.}  
    (\textbf{A}) Example of Rabi oscillations between states $\ket{r_1}$ and $\ket{r_2}$. The sinusoidal fit yields the Rabi frequency of the driving MW field.  
    (\textbf{B}) Linearity of the Rabi frequency versus the set amplitude (a.u.: arbitrary units).  
    (\textbf{C}) Spin-exchange interaction between states $\ket{r_1}$ and $\ket{r_2}$ when the atoms are separated by \SI{37}{\um}. Through fitting the data with a sinusoidal function, we obtain an interaction strength $2\uppi\times\SI{0.316(3)}{MHz}$, corresponding to twice the value of $V$. Error bars denote the binomial standard deviation.  
    (\textbf{D}) Spin-exchange interaction between states $\ket{r_1}$ and $\ket{r_2}$ for atoms separated by \SI{14}{\um}. Using a sinusoidal fitting with a Gaussian-decaying envelope, we find an interaction strength  $2\uppi\times\SI{5.75(4)}{MHz}$ with a $1/\text{e}$ decay time of $\uptau=\SI{0.59(3)}{\us}$. Error bars denote the binomial standard deviation.
  }
  \label{fig:cali_Omega&V}
\end{figure}

\subsection{Calibration of experimental parameters}
\noindent
The driving MW is generated by mixing a local MW source (SMB 100A, Rohde\&Schwarz) with a radio-frequency (RF) signal from an AWG (M4i.6631-x8, Spectrum) using a frequency mixer (ZX05-153LH-S+, Mini-Circuits). The two output channels of the AWG (for state preparation and dissipative evolution) are combined into one via a power splitter (ZFSC-2-1+, Mini-Circuits) before being mixed with the local MW source. The Rabi frequency of the MW, $\Omega$, during the dissipative evolution is calibrated as follows. First, we measure the Rabi oscillation between $\ket{r_1}$ and $\ket{r_2}$ driven by the resonant MW (fig.~\ref*{fig:cali_Omega&V}A) under different amplitudes set to the AWG to verify that the fitted Rabi frequency increases linearly with the set amplitude (fig.~\ref*{fig:cali_Omega&V}B), which means the mixer works in the linear regime. Second, we change the driving MW frequency to resonance with the transition between states $\ket{r_2}$ and $\tfrac{1}{\sqrt{2}}(\ket{r_1}+\ket{e})$ with a large amplitude in the linear regime set to the AWG and measure the PTS dynamics (Fig.~\ref{fig:fig2}C, diamonds) to extract $\Omega$ from the fitting curve with an oscillation frequency $\sqrt{\Omega^2 - \Gamma^2/4}$ and a decay rate $\Gamma/2$. Finally, the linear coefficient between $\Omega$ and the amplitude is determined and we accordingly set the amplitude to control $\Omega$ during the dissipative evolution. Such calibration is free from the problem of non-uniform frequency response of the mixer and other electronic elements.

The dipole-dipole interaction strength between the Rydberg atoms and its fluctuation are determined as follows.
By observing the spin exchange of the two atoms prepared in state $\left(\tfrac{1}{\sqrt{2}}\left(\ket{\uparrow}+\text{i}\ket{\downarrow}\right)\right)\otimes\left(\tfrac{1}{\sqrt{2}}\left(\ket{\uparrow}+\text{i}\ket{\downarrow}\right)\right)$ with the initial expectation value of operator $\upsigma^\text{y}$ equal to $+1$ for each atom (fig.~\ref*{fig:cali_Omega&V}C), the interaction strength $V$ is extracted from the fitting curve. The observed oscillation frequency of $\braket{\upsigma^\text{y}}$ corresponds to the dipole-dipole interaction strength between states $\ket{r_1}$ and $\ket{r_2}$, while that between states $\tfrac{1}{\sqrt{2}}(\ket{r_1}\pm\ket{e})$ and $\ket{r_2}$---denoted as $V$---is half the value. The fluctuation of $V$ is attributed to the motion of the atoms giving rise to uncertainty in $r$, which causes dephasing of the oscillation in the spin exchange dynamics. By fitting the dephasing time $\uptau$, the fluctuation $\updelta V$ is obtained as $\updelta V=\tfrac{\sqrt{2}}{2\uptau}$. A typical example is shown in fig.~\ref*{fig:cali_Omega&V}D, yielding $V=2\uppi \times \SI{2.87(2)}{MHz}$ and $\updelta V=2\uppi\times\SI{0.19(2)}{MHz}$ at an interatomic distance of $r=\SI{14}{\um}$, from which an uncertainty in $r$ of \SI{0.3}{\um} is extracted.
The interaction strengths in Fig.~\ref{fig:fig2}, C to H and Fig.~\ref{fig:fig3} are calculated using $V=\tfrac{1}{2}C_3/r^3$ with $C_3=2\uppi\times\SI{16}{GHz\,\um^3}$ extracted from fig.~\ref*{fig:cali_Omega&V}C.

\subsection{Compensation for light shifts on the dressed state}
\begin{figure}
  \centering
  \includegraphics{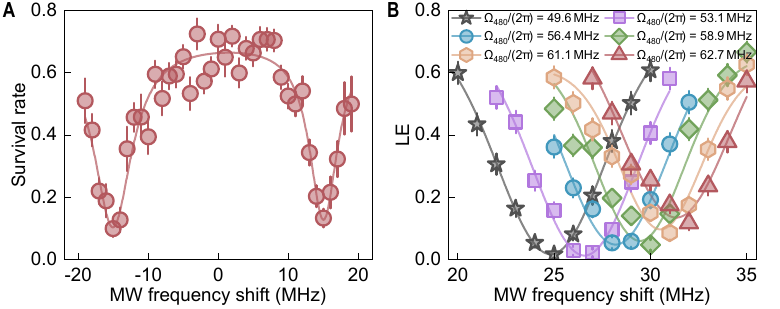}
  \caption{
    \textbf{MW spectra under light shifts.}  
    (\textbf{A}) Typical MW spectra of dressed Rydberg states. The survival probability is measured with atoms prepared in state $\ket{r_2}$ and driven by the MW field under the dressing of the \SI{480}{nm} laser. The data are fitted with a sum of two Lorentzian peaks, separated by the Autler-Townes splitting $\Omega_{480}\approx 2\uppi\times\SI{30}{MHz}$. The spectrum shows a small bias due to the non-uniform frequency response of the MW electronics.  
    (\textbf{B}) MW spectra for different $\Omega_{480}$. The parameters for acquiring the LE after dissipative evolution are $\Omega=2\uppi\times\SI{2.1}{MHz}$, $\Gamma=2\uppi\times\SI{3.0}{MHz}$, and $t=\SI{0.5}{\us}$. In this regime, the LE is expected to approach zero on resonance according to the numerical simulations. The data show that with increasing $\Omega_{480}$, the LE minimum rises, which we attribute to a nonzero real part of the $\upsigma^\text{z}$ term in the Hamiltonian (and a larger maximum imaginary eigenvalue component), induced by fluctuations of the light shift $\Omega_{480}/2$ caused by atomic motions. Error bars denote the binomial standard deviation.
  }
  \label{fig:light_shift}
\end{figure}

\noindent
The \SI{480}{nm} dressing laser induces light shifts on the dressed Rydberg state $\ket{\downarrow}$, causing a nonzero real part of the term $\propto \upsigma^\text{z}$  in $H_\text{pt}$, which breaks its $\mathcal{PT}$ symmetry. Therefore, we need to compensate for this light shift in order to maintain the resonance of the driving MW. We measure the Rabi frequency of the \SI{480}{nm} laser beam $\Omega_{480}$ by MW spectroscopy of the transition from state $\ket{r_2}$ to the dressed state, see fig.~\ref*{fig:light_shift}A. The measured value of $\Omega_{480}$ is used to calibrate the resonant frequency of the driving MW during the dissipative evolution. It is also used for the alignment of the \SI{480}{nm} laser beam, which is controlled by two mirror mounts with piezoelectric adjusters. A good alignment puts the atoms at the center of the laser beam and thus minimizes the influence of the fluctuation of $\Omega_{480}$ induced by the motion of the atoms. The influence remains negligible provided $\Omega_{480}$ does not exceed $2\uppi\times\SI{50}{MHz}$ (fig.~\ref*{fig:light_shift}B). It is estimated that the relative uncertainty of $\Omega_{480}$ is $\sim 5\%$, which is taken into consideration in the numerical simulations. This uncertainty is attributed to the motion of the atoms, corresponding to an axial position spreading of \SI{1.3}{\um} relative to the \SI{480}{nm} laser beam with a Gaussian radius of around \SI{13}{\um}.

\subsection{Tuning the decay rate}
\noindent
We realize a large tuning range of the physical parameters $\Omega/\Gamma$ and $V/\Gamma$ by switching between two different $\ket{e}$ states with distinct spontaneous emission rates. When investigating the $\mathcal{PT}$ phase transition of a two-atom system, the $\ket{e}$ state is chosen to be $\ket{5P_{3/2}}$ which causes the $\ket{\downarrow}$ state to decay at a rate $\Gamma = 2\uppi\times\SI{3.0}{MHz}$. This large decay rate enables a short evolution time which is more than two orders of magnitude shorter than the lifetime of Rydberg states, and therefore keeps the dark count at a low level. On the other hand, when studying the multi-spin chain, state $\ket{5P_{3/2}}$ is replaced by $\ket{6P_{3/2}}$ and reduces the decay rate to $\Gamma = 2\uppi\times\SI{0.7}{MHz}$, which enables the exploration of new phenomena such as the many-body blockade effect under the strong interaction condition $V \gg \Gamma$. The main difference between the two different experimental settings is the dressing laser field during the dissipative stage, switching between \SI{480}{nm} and \SI{1013}{nm} lasers, which couple the Rydberg state to $\ket{5P_{3/2}}$ and $\ket{6P_{3/2}}$, respectively.

While an off-resonant weak dressing (with detuning $\Delta_{480}$ and Rabi frequency $\Omega_{480}$) might seem to allow easier and continuous tuning of $\Gamma$~\cite{begoc2025controlled,chen2025collective}, we prefer the resonant maximal dressing scheme for the following reasons. First, maximal dressing minimizes the uncertainty in the decay rate of the dressed Rydberg state, $\Gamma\propto\uprho_\text{ee}$ (where $\uprho_\text{ee}$ is the population of the intermediate state $\ket{e}$ in the dressed state). This is because variations in $\Gamma$ are of second order with respect to fluctuations in the dressing laser intensity, unlike the off-resonant scheme, where the relationship is of first order. Second, maximal dressing similarly minimizes fluctuations in the interaction strength $V\propto\uprho_\text{rr}$ (where $\uprho_\text{rr}$ is the population of the Rydberg state $\ket{r_1}$ in the dressed state) and in $\Omega\propto\uprho_{rr}^{1/2}$. By suppressing these fluctuations, we significantly reduce the number of Monte-Carlo runs required in numerical simulations to account for variations in $\Omega_{480}$. Finally, the bandwidth of the acoustic-optical modulator (AOM) used to shift the frequency of the dressing laser beam and the available \SI{480}{nm} laser power do not support the set of required parameters where $\Gamma < 2\uppi\times\SI{1}{MHz}$ and $\sqrt{\Omega_{480}^2+\Delta_{480}^2} \gg V$ because when we increase $\Delta_{480}$, $\Omega_{480}$ decreases so rapidly that $\Gamma$ is decreased to around $2\uppi\times\SI{1}{MHz}$ when both $\Delta_{480}$ and $\Omega_{480}$ are smaller than the chosen $V \approx 4\Gamma$.

\section{Theoretical modeling}
\subsection{Single-particle model}
We start with theoretical analysis of the single-particle model to show the PTS--PTSB transition. The eigenvalues of the single-particle $\mathcal{PT}$-symmetric Hamiltonian $H_\text{pt} = \tfrac{\Omega}{2} \upsigma^\text{x} + \text{i}\tfrac{\Gamma}{4}\upsigma^\text{z}$ are $E=\pm \tfrac{1}{2} \sqrt{\Omega^2 - \Gamma^2/4}$. When $\Omega > \tfrac{\Gamma}{2}$, the two eigenvalues are real, and thus $\mathcal{PT}$ symmetry is preserved, where the dynamics (population, coherence, etc) exhibit oscillations at frequency $\sqrt{\Omega^2 - \Gamma^2/4}$. At the exceptional point (EP) with $\Omega = \tfrac{\Gamma}{2}$, both eigenvalues coalesce at zero. When $\Omega < \tfrac{\Gamma}{2}$, the two eigenvalues are both purely imaginary (with opposite signs), and thus $\mathcal{PT}$ symmetry is broken, such that the dynamics show exponential growth in the long term, dominated by the branch with positive imaginary part, while the other decays rapidly.

The Hamiltonian governing the experimental system is $H_\text{nh}=H_\text{pt}+H_\text{im}$, with an additional global decaying term $H_\text{im}=-\text{i}\tfrac{\Gamma}{4}\mathbb{I}$. Therefore, as illustrated in Fig.~\ref{fig:fig2}C, the LE in the PTS phase is a damped oscillation decaying at rate $\Gamma/2$, while in the PTSB regime a slower decay rate is observed.

Here, we prove the $\mathcal{PT}$ symmetry, $[\mathcal{PT},H_\text{pt}]=0$. Here $\mathcal{T}$ is the conjugate operator ($\mathcal{T}\rm{A}=\rm{A}^*$ for any operator A). In the matrix form, the parity operator reads,
\begin{equation*}
    \mathcal{P}=
    \left(
    \begin{array}{cc}
    0 & 1\\
    1 & 0
    \end{array}
    \right),
\end{equation*}
and the Hamiltonian is,
\begin{equation*}
    H_\text{pt}=
    \left(
    \begin{array}{cc}
    \text{i}\Gamma/4 & \Omega/2\\
        \Omega/2 & -\text{i}\Gamma/4
    \end{array}
    \right).
\end{equation*}
We have $\mathcal{T}^{-1}=\mathcal{T}$ and $\mathcal{P}^{-1}=\mathcal{P}$, and therefore one finds that,
\begin{align*}
    \mathcal{PT}H_\text{pt}(\mathcal{PT})^{-1}
    &=\mathcal{PT}H_\text{pt}\mathcal{T}^{-1}\mathcal{P}^{-1}=\mathcal{P}\mathcal{T}H_\text{pt}\mathcal{T}\mathcal{P}=\mathcal{P}(\mathcal{T}H_\text{pt})(\mathcal{T}\mathcal{P})\\
    &=
    \left(
    \begin{array}{cc}
    0 & 1\\
    1 & 0
    \end{array}
    \right)
    \left(
    \begin{array}{cc}
    -\text{i}\Gamma/4 & \Omega/2\\
        \Omega/2 & \text{i}\Gamma/4
    \end{array}
    \right)
    \left(
    \begin{array}{cc}
    0 & 1\\
    1 & 0
    \end{array}
    \right)\\
    &=
    \left(
    \begin{array}{cc}
    \text{i}\Gamma/4 & \Omega/2\\
        \Omega/2 & -\text{i}\Gamma/4
    \end{array}
    \right)\\
    &=H_\text{pt}.
\end{align*}
This allows us to write down the commutation relation, $[\mathcal{PT},H_\text{pt}]=0$. 
Hence Hamiltonian $H_{\text{pt}}$ preserves the $\mathcal{PT}$ symmetry. 
\begin{figure}
  \centering
  \includegraphics{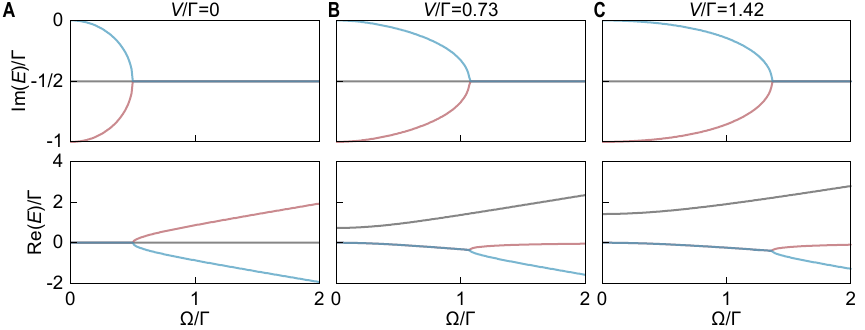}
  \caption{
    \textbf{Eigenvalues of the two-atom Hamiltonian.}  
    Imaginary and real parts of the eigenvalues of the Hamiltonian Eq.~\ref{eq:32}, for interaction strengths $V=0$ (\textbf{A}), $V=0.73\Gamma$ (\textbf{B}), and $V=1.42\Gamma$ (\textbf{C}). The parameters are those used in Fig.~\ref{fig:fig2}, A and B.
  }
  \label{fig:eigen}
\end{figure}

\subsection{Interacting Rydberg atoms model}
\label{secSM:twoatom}
\noindent
The Hamiltonian governing the interacting two-atom system, $H = \sum_i \left( \tfrac{\Omega}{2} \upsigma^\text{x}_i + \text{i} \tfrac{\Gamma}{4} \upsigma^\text{z}_i -\text{i}\tfrac{\Gamma}{4}\mathbb{I}_i\right) + \sum_{i<j} V_{ij} \left( \upsigma^+_i \upsigma^-_j + \upsigma^+_j \upsigma^-_i \right)$, is rewritten in the matrix form as,
\begin{equation}
    H=
    \left(
    \begin{array}{cccc}
        0 & \frac{\Omega}{2} & \frac{\Omega}{2} & 0 \\
        \frac{\Omega}{2} & -\text{i}\frac{\Gamma}{2} & V & \frac{\Omega}{2} \\
        \frac{\Omega}{2} & V & -\text{i}\frac{\Gamma}{2} & \frac{\Omega}{2} \\
        0 & \frac{\Omega}{2} & \frac{\Omega}{2} & -\text{i}\Gamma
    \end{array}
    \right),
\end{equation}
in the atomic basis $\{ \ket{\uparrow\uparrow}, \ket{\uparrow\downarrow}, \ket{\downarrow\uparrow}, \ket{\downarrow\downarrow} \}$.
Since the anti-symmetric state $\tfrac{1}{\sqrt{2}}(\ket{\downarrow\uparrow}-\ket{\uparrow\downarrow})$ is a dark state, the Hamiltonian is simplified as
\begin{equation}
    H=
    \left(
    \begin{array}{ccc}
        0 & \frac{\Omega}{\sqrt{2}} & 0 \\
        \frac{\Omega}{\sqrt{2}} & V-\text{i}\frac{\Gamma}{2} & \frac{\Omega}{\sqrt{2}} \\
        0 & \frac{\Omega}{\sqrt{2}} & -\text{i}\Gamma
    \end{array}
    \right),
    \label{eq:32}
\end{equation}
in the basis $\{ \ket{\uparrow\uparrow}, \tfrac{1}{\sqrt{2}}(\ket{\uparrow\downarrow}+\ket{\downarrow\uparrow}), \ket{\downarrow\downarrow} \}$.
The dynamics can be readily obtained numerically. Representative results are shown in fig.~\ref*{fig:eigen}. Under the strong-interaction condition $V \gg \Omega,\Gamma$, the single-excitation state $\tfrac{1}{\sqrt{2}}(\ket{\uparrow\downarrow}+\ket{\downarrow\uparrow})$ is adiabatically eliminated. The Hamiltonian is simplified to be
\begin{equation}
    H=
    \left(
    \begin{array}{cc}
        0 & \Omega_\text{eff}/2\\
        \Omega_\text{eff}/2 & -\text{i}\Gamma_\text{eff}/2
    \end{array}
    \right),
\end{equation}
with $\Omega_\text{eff}=\tfrac{\Omega^2V}{V^2+\Gamma^2/4}$ and $\Gamma_\text{eff}=2\Gamma(1+\tfrac{\Omega^2/4}{V^2+\Gamma^2/4})$. This allows us to find the EP approximately,
\begin{equation}
    \Omega=\sqrt{\frac{V^2+\Gamma^2/4}{|V|-\Gamma/4}\Gamma}\approx\frac{\Gamma}{2}\sqrt{1+4\frac{|V|}{\Gamma}}.
\end{equation}
Remarkably, the EP predicted by the strong-interaction approximation agrees with numerical solutions even at $V=0$.

\subsection{Quantum master equation and non-Hermitian Hamiltonian formalism}
\noindent
In light-atom interacting systems, dynamics of the electronic states are usually described by the Lindblad master equation. In our experiment, two Rydberg states $\ket{r_1} = \ket{68D_{5/2}}$ and $\ket{r_2} = \ket{69P_{5/2}}$, as well as the two low-lying states $\ket{e}=\ket{5P_{3/2}}$ and $\ket{g}=\ket{5S_{1/2}}$ are involved.
The master equation is given by
\begin{equation}
    \frac{\partial}{\partial t} \uprho = - {\rm i} \left[ H, \uprho \right] + \sum_i \Gamma \mathcal{D}[ \ket{g_i}\bra{e_i}] \uprho,
    \label{seq:ME}
\end{equation}
where $\mathcal{D}[\hat{o}] \uprho = \hat{o} \uprho \hat{o}^\dagger - \tfrac{1}{2} \left( \hat{o}^\dagger \hat{o} \uprho + \uprho \hat{o}^\dagger \hat{o} \right)$. In the rotating frame, the Hamiltonian reads
\begin{align}
    H &= \sum_{i} \frac{\Omega_{480}}{2} \left( \ket{e_i}\bra{r_{2,j}} + \ket{r_{2,j}}\bra{e_i} \right) + \frac{\Omega_{\upmu \rm w}}{2} \left( \ket{r_{1,j}}\bra{r_{2,j}} + \ket{r_{2,j}}\bra{r_{1,j}} \right) + \Delta \ket{r_{2,j}}\bra{r_{2,j}} \nonumber \\
    &\phantom{={}} + \sum_{i < j} V_{\rm dd}(\mathbf{r}_{i}-\mathbf{r}_{k}) \left( \ket{r_{1,j}r_{2,k}} \bra{r_{2,j}r_{1,k}} + \ket{r_{2,j}r_{1,k}} \bra{r_{1,j}r_{2,k}} \right).
\end{align}
The MW field resonantly couples states $\ket{r_2}$ and $\ket{\downarrow} = (\ket{r_1} + \ket{e})/\sqrt{2}$. The qubit states are encoded in states $\ket{\uparrow} = \ket{r_2}$ and $\ket{\downarrow}$. State $\ket{d} = (\ket{r_1} - \ket{e})/\sqrt{2}$ is effectively decoupled due to the large detuning $\Delta'=\Omega_{480}$. In the qubit basis, the Hamiltonian is $H = H_0 + H_\text{d}$, where 
\begin{align}
H_0 &= \sum_i \frac{\Omega}{2} \left( \ket{\uparrow_i}\bra{\downarrow_i} + \ket{\downarrow_i}\bra{\uparrow_i} \right) + \sum_{i<j} V_{ij} \left( \ket{\uparrow_i\downarrow_j}\bra{\downarrow_i\uparrow_j} + \ket{\downarrow_i\uparrow_j}\bra{\uparrow_i\downarrow_j} \right), \\
    H_\text{d} &= \sum_i \Delta' \ket{d_i}\bra{d_i} + \frac{\Omega}{2} \left( \ket{\uparrow_i}\bra{d_i} + \ket{d_i}\bra{\uparrow_i} \right) + \sum_{i<j} V_{ij} \left( \ket{\uparrow_id_j}\bra{d_i\uparrow_j} + \ket{d_i\uparrow_j}\bra{\uparrow_id_j} \right),
\end{align}
where $\Omega = \Omega_{\upmu \rm w}/\sqrt{2}$ and $V_{ij} = V_{\rm dd}(\mathbf{r}_i - \mathbf{r}_j)/2 \propto |j-k|^{-3}$.
States $\ket{\downarrow}$ and $\ket{d}$ decay to the ground state with rate $\Gamma = \upgamma/2$. 

In our system, state $\ket{d}$ is adiabatically eliminated as it is energetically far-detuned. We obtain the $N$-atom Hilbert space $\mathscr{H} = \{ \ket{\uparrow}, \ket{\downarrow}, \ket{g} \}^{\otimes N}$, and the space $\mathscr{Q} = \{ \ket{\uparrow}, \ket{\downarrow} \}^{\otimes N}$ which excludes $\ket{g}$ state. Since there is no coherent coupling between subspaces $\mathscr{Q}$ and $\mathscr{H}\backslash \mathscr{Q}$, the density matrix is decomposed to the form,
\begin{equation}
    \uprho = \uprho_\mathscr{Q} \oplus \uprho_{\mathscr{H}\backslash\mathscr{Q}}.
\end{equation}
Dynamics of the space $\mathscr{Q}$ is governed by the master equation,
\begin{equation}
    \frac{\partial}{\partial t} \uprho_\mathscr{Q} = -{\rm i} \left[ H_0, \uprho_\mathscr{Q} \right] - \sum_i \frac{\Gamma}{2} \left( \ket{\downarrow_i}\bra{\downarrow_i} \uprho_\mathscr{Q} + \uprho_\mathscr{Q} \ket{\downarrow_i}\bra{\downarrow_i} \right),
\end{equation}
which allows us to decompose the density matrix with the unnormalized wave function, $\uprho_\mathscr{Q} = \ket{\uppsi}\bra{\uppsi}$. We then obtain the non-Hermitian Schr\"odinger equation
\begin{equation}
    {\rm i} \frac{\partial}{\partial t} \ket\uppsi = \left(H_0 - {\rm i} \frac{\Gamma}{2} \sum_i \ket{\downarrow_i}\bra{\downarrow_i} \right) \ket\uppsi,
    \label{eq:nHSE}
\end{equation}
which has been used in the main text. It is one of the key results for exploring non-Hermitian many-body physics using the Rydberg atom array. 

\begin{figure}
  \centering
  \includegraphics{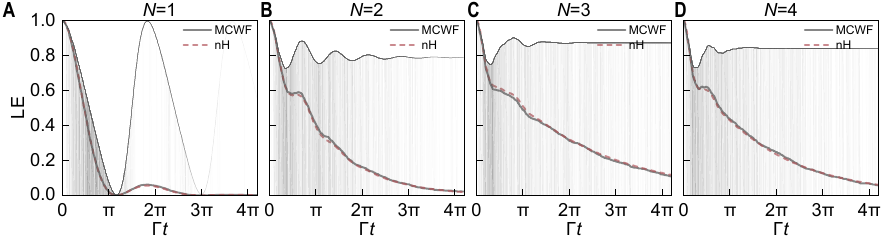}
  \caption{
    \textbf{Comparison between master-equation and non-Hermitian approaches.}  
    Thin gray lines show 1000 realizations of Monte Carlo wave functions (MCWFs); their average is indicated by the bold gray line. The dashed red line gives the result from numerically solving the non-Hermitian (nH) Schr\"odinger equation. Results are shown for $N=1$--$4$ in (\textbf{A} to \textbf{D}), respectively, with parameters $\Omega=1.2\Gamma$ and $V=2.0\Gamma$.
  }
  \label{fig:ME}
\end{figure}

To confirm the validity of the non-Hermitian model (fig.~\ref*{fig:ME}, red dashed lines), we solve the master equation (Eq.~\ref{seq:ME}, gray solid lines) numerically by unraveling it through the Monte Carlo wave function (MCWF) approach~\cite{ashidaNonHermitianPhysics2020}. The MCWF approach is one of the standard methods to solve the master equation. Starting from the initial state, one evolves the many-body wave function with quantum jumps that occur randomly at each time step. Many stochastic trajectories (MCWFs) are generated, with which expectation values of  operators can be calculated as the ensemble average. In fig.~\ref*{fig:ME}, we show dynamics of the LE obtained from the MCWF and non-Hermitian Hamiltonian situations. In the MCWF simulation, we have realized over 1000 trajectories with the given parameters. The averaged LE from the MCWF simulation shows almost identical results to the non-Hermitian calculation. The minor differences result from the finite number of the trajectories. 

While our analysis in the main text has been presented in terms of the Schrödinger equation (SE) with an effective non-Hermitian Hamiltonian, this description is fully equivalent to the master-equation (ME) formalism in the regime of interest. Specifically, the non-Hermitian term in the SE arises from tracing over quantum jump processes in the ME and restricting to the no-jump subspace, which exactly captures the dynamics of LE---an observable that becomes zero if any quantum jump occurs---making the no-jump description directly relevant. The SE formalism thus provides a compact and transparent representation of the same physical dynamics encoded in the full ME, while avoiding the additional computational overhead of tracking the stochastic jump trajectories or the full density matrix. This simplification allows us to highlight the essential interplay of coherent driving, dissipation, and interactions responsible for the novel effects observed in our experiment, without loss of generality and accuracy in interpreting the results.

\section{Dynamics of the non-Hermitian many-body $XY$ model}
\noindent
Diagonalizing the non-Hermitian Hamiltonian $H$, the right and left eigenvectors, $\ket{v_k}$ and $\bra{w_k}$, can be obtained. They satisfy the eigenvalue equations, $H \ket{v_k} = E_k \ket{v_k}$ and $\bra{w_k} H = E_k \bra{w_k}$, with eigenvalue $E_k$. The basis $\ket{v_k}$ and $\bra{w_k}$ form the biorthonormal condition, 
\begin{equation}
	\sum_k \ket{v_k}\bra{w_k} = \mathbb{I}. 
\end{equation}
Depending on the symmetry of the Hamiltonian, the spectrum can be real-valued or complex. With the eigenvalues and eigenvectors, we evaluate state $\ket{\uppsi_N(t)}$ at time $t$ via $\ket{\uppsi_N(t)} = \sum_k\Braket{w_k|0_N}\text{e}^{-\text{i}E_kt}\ket{v_k}$ with initial state $\ket{0_N}$.

\subsection{Loschmidt Echo}
\noindent
The strong correlation between the Loschmidt Echo (LE) and the maximum imaginary part of the eigenvalues arises because the system evolves predominantly into the eigenstate that decays most slowly when the evolution time is sufficiently long. This is explained by the following analysis of the Hamiltonian and its eigenvalues of this non-Hermitian system.
The LE quantifies the survival probability of the initial state $\ket{0_N}$, 
\begin{align}
    F &= \left| \bra{0_N} \text{e}^{-\text{i} H t} \ket{0_N} \right|^2 \nonumber\\
    &= \left| \sum_k \text{e}^{-\text{i} E_k t} \Braket{0_N|v_k} \Braket{w_k|0_N} \right|^2 \nonumber \\
    &=\sum_k\text{e}^{\text{i}(E_k^*-E_k)t}\left|\braket{w_k|0_N}\right|^2\left|\braket{0_N|v_k}\right|^2 + \sum_{ij}\text{e}^{\text{i}(E_j^*-E_i)t}\braket{0_N|v_i}\braket{v_j|0_N}\braket{w_i|0_N}\braket{0_N|w_j},
    \label{eqSM:LE}
\end{align} 
where $\braket{0_N|w_k}$ ($\braket{0_N|v_k}$) is the projection of the initial state to eigenstate $\ket{w_k}$ ($\ket{v_k}$).
The first term in Eq.~\ref{eqSM:LE} gives the diagonal coupling of the eigenstate, while the second term is the off-diagonal coupling of between different eigenstates. When the Hamiltonian is $\mathcal{PT}$-symmetric and the entire spectrum is real-valued, the diagonal coupling becomes time-independent, while the off-diagonal coupling terms oscillates in time. In this case, the LE is bounded by 1. If the $\mathcal{PT}$ symmetry is broken, exponentially growing terms will appear in the LE due to complex $E_k$. However, this will not happen in our experiment, as dynamics of the atoms are governed by Hamiltonian $H_\text{nh}=H_\text{pt}-\text{i} \tfrac{N\Gamma}{4}\mathbb{I}$. This gives a global decay term, $\text{e}^{-N\Gamma t/2}$, to $F(t)$. In the PTS phase, the oscillation amplitude damps with time when $t\to \infty$. In the PTSB phase, the exponentially growing term, on the other hand, counteracts the decay in the long-term evolution. 
We use the rate function $\uplambda(t) = -\ln F(t)$ to capture the different exponential decay behaviors of the LE in PTS and PTSB phases. Since the oscillation in the PTS phase is bounded, we can apply a single time measurement to distinguish the PTS and PTSB phases within the tolerance that $| {\rm Im}({E_k}) | < \upepsilon$ for all $E_k$. The minimal time is approximately given by,
\begin{equation}
    t_{\min} \sim \frac{\ln 2}{\upepsilon}.
\end{equation}

\begin{figure}
  \centering
  \includegraphics[width=\textwidth]{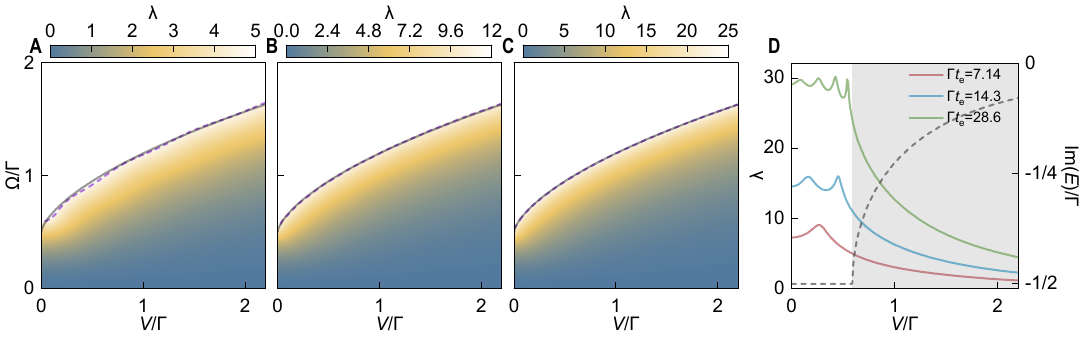}
  \caption{
    \textbf{$\mathcal{PT}$ phase diagram for two Rydberg atoms at various evolution times.}  
    Rate function $\uplambda(t_\text{e})$ at $\Gamma t_\text{e}=$ 7.14 (\textbf{A}), 14.3 (\textbf{B}), and 28.6 (\textbf{C}), shown in the color scale. Gray solid curves mark the EPs obtained from eigenvalue analysis of the Hamiltonian (see Sec.~\ref{secSM:twoatom}), while purple dashed lines are contours at $\uplambda=4.7$, 10.8, and 23.7, respectively.  
    (\textbf{D}) Maximal imaginary part of the eigenvalues Im$(E)$ (gray dashed line, right axis) and the rate function $\uplambda(t_\text{e})$ (colored solid lines, left axis). The shaded region denotes the PTSB phase, where $\uplambda$ decreases with increasing $V$, while in the PTS phase $\uplambda$ exhibits oscillatory behavior. Simulations use a fixed $\Omega/\Gamma=1$.
  }
  \label{fig:twobody}
\end{figure}

In fig.~\ref*{fig:twobody}, A to C, rate functions at different time $t_{\text{e}}$ are shown. Fixing $V$, one can see that the rate decreases when increasing $\Omega$. We define a critical $\uplambda$ as the boundary between the PTS and PTSB phase. This boundary is compared with the one obtained from analyzing the eigenvalues of the Hamiltonian discussed in Sec.~\ref{secSM:twoatom}. When $t_{\text{e}}$ is small (fig.~\ref*{fig:twobody}A), the rate function prediction deviates slightly from the one from the eigenvalues. The cause of the minor discrepancy arises from the oscillations of the LE in the regime where $\Omega \gg V$. Increasing the evolution time, the oscillation decays. As a result, the phase boundary obtained from the two methods agrees well (see panels B and C). Figure~\ref*{fig:twobody}D elucidates how the LE characterizes the $\mathcal{PT}$ phase transitions: Within the PTSB regime, the rate function decreases as the maximum imaginary component of eigenvalues increases, while in the PTS regime, oscillatory behaviors of the rate function depend on $V$. In practice, a larger $t_{\text{e}}$ (or equivalently, $\uplambda$) is required to determine the phase boundary reliably.

\begin{figure}
  \centering
  \includegraphics[width=\textwidth]{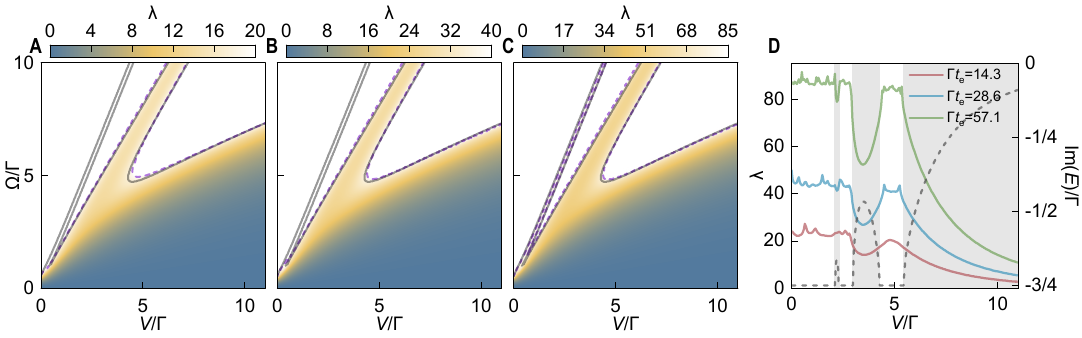}
  \caption{
    \textbf{$\mathcal{PT}$ phase diagram for three Rydberg atoms at various evolution times.}  
    Rate function $\uplambda(t_\text{e})$ at $\Gamma t_\text{e}=$ 14.3 (\textbf{A}), 28.6 (\textbf{B}), and 57.1 (\textbf{C}), shown in the color scale. Gray solid curves mark the EPs obtained from eigenvalue analysis of the Hamiltonian, while purple dashed lines are contours at $\uplambda=19.5$, 39.5, and 81.7, respectively.
    (\textbf{D}) Maximal imaginary part of the eigenvalues Im$(E)$ (gray dashed line, right axis) and the rate function $\uplambda(t_\text{e})$ (colored solid lines, left axis). The shaded region denotes the PTSB phase, where $\uplambda$ decreases with increasing $V$. In the PTS phase $\uplambda$ exhibits oscillatory behavior. In the simulations we fix $\Omega/\Gamma=5$.
  }
  \label{fig:threebody}
\end{figure}

Increasing the system size $N$, the required evolution time is much longer to distinguish between the PTS and PTSB phases based on the LE as the phase diagram becomes more complex with more branches of the PTSB phase~\cite{lourencco2022non}. As shown in fig.~\ref*{fig:threebody}, A and B, insufficient evolution time may impede the resolution of fine-grained bifurcation patterns in the $\mathcal{PT}$ phase transition of the three-atom case. A selected evolution time to achieve marginal distinguishability between PTS and PTSB phases is $\Gamma t_\text{e}=57.1$ (fig.~\ref*{fig:threebody}C), corresponding to $\uplambda(t_\text{e})=85$, which presents substantial experimental challenges. The capability of the LE as a $\mathcal{PT}$ phase indicator is further demonstrated in fig.~\ref*{fig:threebody}D, where both the amplitudes and the oscillatory characteristics exhibit strong correlations with $\mathcal{PT}$ phase transitions.

\begin{figure}
  \centering
  \includegraphics{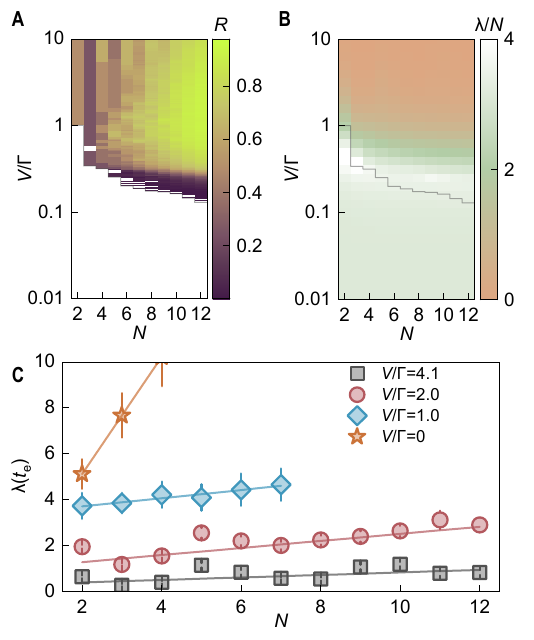}
  \caption{
    \textbf{Multi-atom $\mathcal{PT}$ phase diagram.}  
    (\textbf{A}) Contour plot of the participation ratio $R$, corresponding to Fig.~\ref{fig:fig4}A. The gray line delineates the operational phase boundary, defined by connecting local minima of transition points along the $N$-axis. The exact phase boundary between zero and nonzero values is indicated by the transition from white to purple.  
    (\textbf{B}) Contour plot of the averaged rate function $\uplambda/N$ at $\Omega/\Gamma=1.2$ and evolution time $\Gamma t_\text{e}=7.14$. The gray line is the same phase boundary as in (A). The color scale captures the phase boundary only approximately, as the evolution time is not sufficiently long.  
    (\textbf{C}) Rate function $\uplambda(t_\text{e})$ obtained from the experimental data in Fig.~\ref{fig:fig4}B. Solid lines show linear fits, with larger slopes for smaller $V/\Gamma$. While the fits agree with data points in the PTS phase ($V/\Gamma=0$), they become increasingly inaccurate in the PTSB phase ($V/\Gamma>1.0$) due to the non-Hermitian many-body blockade effect.
  }
  \label{fig:Imfrac}
\end{figure}

Further increasing the system size $N$, there are more tiny PTSB branches, as shown in the contour plot of the participation ratio $R$, i.e., the fraction of eigenvalues that are imaginary (fig.~\ref*{fig:Imfrac}A). Larger $\uplambda$ is needed to distinguish these branches as the system size increases, because the decay rate of the LE is multiplied by $N$. The balanced rate function, defined by $\uplambda/N$, is able to characterize the phase transition (fig.~\ref*{fig:Imfrac}B). An approximate phase boundary is  traced by the contour plot of $\uplambda/N$. The PTSB area with a relatively small value of $\uplambda/N$ (light green) largely overlaps with the dark purple area in fig.~\ref*{fig:Imfrac}A. This results from the fact that the initial state exhibits a small projection onto the eigenstate whose eigenenergy has the largest imaginary part in this area. Hence it takes longer evolution time to distinguish this branch that decays slightly slower than the PTS phase. A clear distinction between the two phases requires evolving the system under dissipation for sufficiently long times, at which point the LE decays to values too small to be faithfully captured within the numerical accuracy of our simulations.

\subsection{Inverse participation ratio}
\label{secSM:IPR}
\noindent
The dependence of the LE on $N$ shows that the initial state will predominantly couple to a few spin-wave states in the singly excitation sector. The ``localization'' of the wave function can be quantified by the inverse participation ratio (IPR)~\cite{lezamaEquilibrationTimeManybody2021}, 
\begin{equation}
    \mathcal{I}(t)=\sum_k \left|\braket{v_k|\uppsi_N(t)}\right|^4,
\end{equation}
where $\ket{v_k}$ is some many-body basis of Hamiltonian $H$. It thus measures the spreading of the initial state $\ket{0_N}$ over the many-body space. Larger (smaller) IPRs indicate that an initial state couples to many (few) different many-body states. In Hermitian systems, IPR strongly connects to the LE as well as R\'enyi entropy, and is widely used as an indicator of quantum chaos. Projecting to many-body basis $\ket{v_k}$, $\mathcal{I}$ is given by, 
\begin{equation}
    \mathcal{I}(t)=\sum_k \text{e}^{2\text{i}(E_k^*-E_k)}\left|\braket{w_k|0_N}\right|^2\left|\braket{0_N|v_k}\right|^2.
\end{equation}
At longer times, only slow decaying modes contribute to $\mathcal{I}(t)$. This is particularly important in the PTSB phase, where the initial state will predominantly couple to a few spin-wave states in the singly excitation sector. Note that the many-body blockade effect is largely observed in the PTSB phase, and the IPR quantifies the ``localization'' of the wave function due to the blockade.

It is interesting to examine the LE in this regime. The off-diagonal coupling mixes the slow and fast decay modes. At later times, the corresponding coupling makes less contribution to dynamics of the LE. Hence one could approximate LE as, 
\begin{equation}
    F(t) \approx \sum_k\text{e}^{\text{i}(E_k^*-E_k)t}\left|\braket{w_k|0_N}\right|^2\left|\braket{0_N|v_k}\right|^2.
\end{equation}
If only a few slow decay modes dominate, or eigenvalues of these modes have same (similar) imaginary parts, $\mathcal{I}(t)$ and $F(t)$ would depend similarly on these slow modes, while $\mathcal{I}(t)$ decays faster than $F(t)$ (the rate differs by a factor of $2$). This is generally true for different system size $N$ when the dynamics are stable (e.g., $\Gamma t=6\uppi$ in fig.~\ref*{fig:norLE}A), as shown in fig.~\ref*{fig:norLE}B. The correlation between the IPR and LE suggests a way to estimate IPR through measuring the LE.

\begin{figure}
  \centering
  \includegraphics{Figure_SI/fig_norLE.pdf}
  \caption{
    \textbf{Correlation between normalized LE and IPR.}  
    (\textbf{A}) Numerical simulation of the dynamical evolution of the normalized LE, $\tilde{F}(t)$, for different system sizes with parameters $\Omega=1.2\Gamma$ and $V=4.1\Gamma$.  
    (\textbf{B}) Black (left axis): $\tilde{F}(t_\text{s})$ at evolution time $\Gamma t_\text{s}=6\uppi$ when the system reaches a quasi-stationary state, using the same values as in (A). Red (right axis): saturated inverse participation ratio (IPR), $\mathcal{I}(t_\text{s})$, in the computational basis.
  }
  \label{fig:norLE}
\end{figure}

\subsection{Quantum Zeno effect}
\begin{figure}
    \centering
    \includegraphics{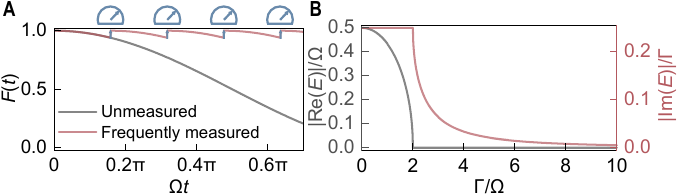}
    \caption{
    \textbf{Illustration of the quantum Zeno effect.}  
    (\textbf{A}) LE dynamics of the early stage of a Rabi oscillation with and without frequent measurements. The measurements project the state onto the initial state with high probability, effectively ``freezing'' the LE near 1.  
    (\textbf{B}) Real and imaginary parts of the single-atom non-Hermitian Hamiltonian as a function of dissipation (measurement) strength $\Gamma$.
    }
    \label{fig:zeno}
\end{figure}

\noindent
We offer an alternative perspective on dissipation, interpreting it as a continuous projection measurement of quantum states, with leakage into the environment. As the dissipation rate $\Gamma$ increases, the measurement action becomes strong, resulting in a more significant quantum Zeno effect (QZE)~\cite{itano1990quantum}. This continuous measurement slows down the dynamics of the system that is driven coherently (fig.~\ref*{fig:zeno}A). This is seen in the decrease of the LE's oscillation frequency, $2\left|\text{Re}(E)\right|=\sqrt{\Omega^2-\Gamma^2/4}$, for a single atom evolving in the PTS regime, as the decay rate $\Gamma$ increases (fig.~\ref*{fig:zeno}B, gray curve). The oscillation becomes ``frozen'' when the dissipation (measurement) is sufficiently strong to drive the system into the PTSB regime, where $\left|\text{Re}(E)\right|=0$. Further increasing $\Gamma$, even the decaying behavior is slowed down counterintuitively as the stabilized decay rate $2\left|\max{\text{Im}(E)}\right|=\Gamma/2-\sqrt{\Gamma^2/4-\Omega^2}$ becomes smaller, signifying the strengthening of the QZE (fig.~\ref*{fig:zeno}B, red curve).

For a strongly interacting multi-atom system, the interplay of interactions and the QZE induced by dissipation creates intriguing phenomena. The interactions slow down the evolution of the system. This significantly enhances the QZE, accompanied by the interaction-driven PTS--PTSB phase transitions. 
The QZE also blockades the system from being excited to multi-excitation manifolds more effectively, since the decay rate of multi-excitation states is multiplied by the excitation number, i.e., the ``measurement'' frequency. 
The QZE confines the multi-atom system to the single-excitation subspace and facilitates the observation of the non-Hermitian many-body blockade effect which exhibits significant dependence on the system size, where the single-excitation subspace that forms collective spin-wave states plays an important role.

\subsection{Spin-wave analysis}

\begin{figure}
    \centering
    \includegraphics[width=\textwidth]{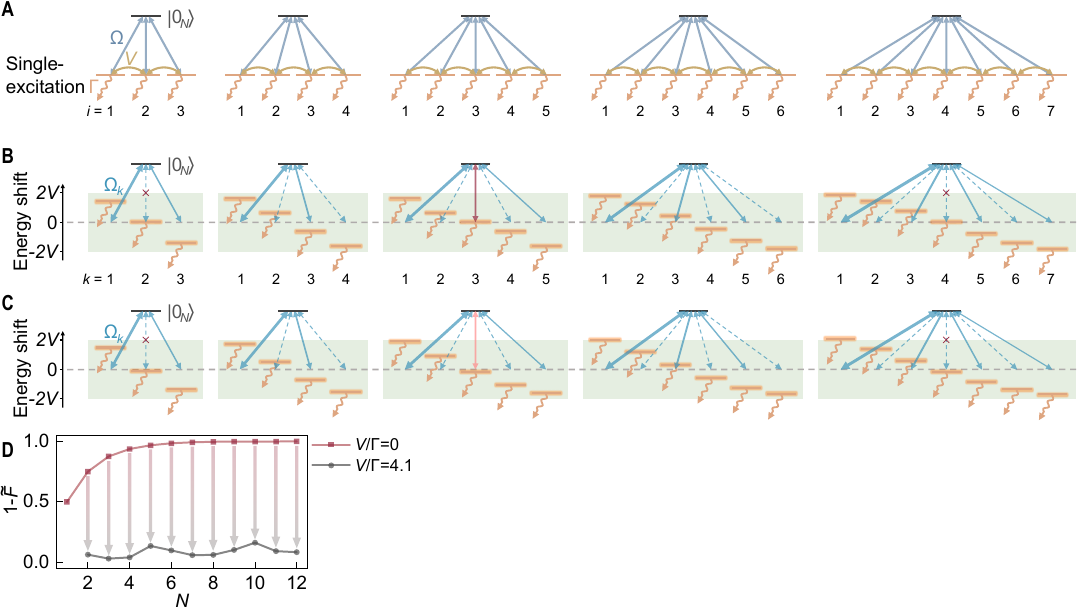}
    \caption{
    \textbf{Multi-atom spin-wave model.}  
    Energy level diagrams for $N=3$--7:  
    (\textbf{A}) Computational basis states in the single-excitation manifold (green lines) with decay rate $\Gamma$, $\upsigma_i^\text{x}\ket{0_N}=\ket{\downarrow\uparrow\uparrow\cdots}, \ket{\uparrow\downarrow\uparrow\cdots}, \dots, \ket{\cdots\uparrow\uparrow\downarrow}$. States are coupled to each other by the NN interaction $V$ and to the initial state $\ket{0_N}$ (black line) via resonant driving with Rabi frequency $\Omega$ (blue bi-directional arrows).  
    (\textbf{B}) Diagonalized spin-wave states (orange lines) $k=1,2,\dots,N$ under NN interactions, with individual energy shifts $U_k$ and couplings $\Omega_k$ to $\ket{0_N}$.  
    (\textbf{C}) Diagonalized spin-wave states under full-range interaction, with energy shifts $U_k$ and couplings $\Omega_k$ to $\ket{0_N}$. In (B) and (C), $U_k$ are referenced to the left axis, with green shaded areas indicating the range $[-2V,2V]$. Couplings $\Omega_k$ are represented by the thickness of the bi-directional arrows (nonlinearly scaled), with dashed arrows indicating zero coupling for anti-symmetric spin-wave states.  
    (\textbf{D}) Comparison of normalized excitation probability $1-\tilde{F}(t\to\infty)$ between non-interacting ($V=0$) and strongly interacting ($V=4.1\Gamma$) cases. For the non-interacting case, an average excitation probability of 0.5 per atom yields $\tilde{F}=0.5^N$.
    }
    \label{fig:spinwave}
\end{figure}

\noindent
In the multi-atom regime, we investigate many-body non-Hermitian physics under the single-excitation approximation as facilitated by the QZE. This approximation is supported by numerical simulations in Fig.~\ref{fig:fig5}E: the single-excitation probabilities are significant, while contributions from double and higher excitations are negligible.

Inspired by the numerical simulation, we analyze the elementary excitations of the system using an effective non-Hermitian Hamiltonian, $H_{\text{eff}} = \sum_k \left( U_k + \text{i}\Gamma \right) \left( \ket{0_N} \bra{0_N} - \ket{k} \bra{k} \right) + \Omega_k \left( \ket{0_N} \bra{k} + \ket{k} \bra{0_N} \right)$, where $\ket{k}$ is the spin-wave state in the single-excitation manifold with 1 atom at $\ket{\downarrow}$ state. As illustrated in fig.~\ref*{fig:spinwave}A, the ground state $\ket{0_N}=\ket{\uparrow}^{\otimes N}$ is coupled to the single excited states in the computational basis $\{\ket{1_i}\}=\{ \ket{\downarrow\uparrow\uparrow\cdots}, \ket{\uparrow\downarrow\uparrow\cdots}, \dots, \ket{\cdots\uparrow\uparrow\downarrow} \}$ with equal strengths and phases. These states are coupled to each other by the dipolar exchange interaction. We are able to analytically evaluate the energy shift $U_k = 2V \cos\left(\tfrac{k\uppi}{N+1}\right)$ and coupling strength to the ground state $\Omega_k = \Omega \tfrac{\sum_i \sin\left[ik\uppi/(N+1)\right]}{\sqrt{ \sum_i \sin^2 \left[ik\uppi/(N+1)\right]}}= \Omega\left[1-(-1)^k\right] \sqrt{\tfrac{2}{N+1}} \cot\left[\tfrac{k \uppi}{2(N+1)}\right]$ by taking into account only the nearest neighbor (NN) interactions (fig.~\ref*{fig:spinwave}B and Fig.~\ref{fig:fig5}B).

With the full dipolar interaction, the zero-energy state for odd $N$ becomes detuned from resonance (fig.~\ref*{fig:spinwave}C), thus suppressing resonant coupling channels and causing additional blockade effects for $N=5,9,\dots$. The anti-symmetric forbidden channels in the NN model---which exhibit the strongest blockade effects for $N=3,7,\dots$---remain prohibited under the full interaction, as the symmetry remains unchanged when long-range interactions are included. This dichotomy explains the differences in LE between NN and full-range interaction models (Fig.~\ref{fig:fig4}B), as the long-range interactions enhance the LE for $N=5,9,\dots$ through supplementary blockade mechanisms, while preserving the strong blockade effect for $N=3,7,\dots$ predicted by the NN interaction model. 

On the other hand, the blockade effect for even $N$ is also modified by the long-range interactions, with the two middle states shifted in the same direction. For $N=4,8,\dots$, the coupled state is shifted away from resonance, thus enhancing the blockade effect. For $N=6,10,\dots$, the coupled state moves closer to the resonance, thus reducing the blockade effect. As the energy shifts induced by the long-range interaction become more significant with increasing $N$, the nearly resonant state can move rightwards as shown in fig.~\ref*{fig:spinwave}C, periodically modifying the blockade strength, which de-synchronizes the dynamical behavior from the ideal period-4 oscillation predicted by the NN model.

We further characterize the strength of the non-Hermitian many-body blockade using the suppression of the excitation probability for different system sizes (fig.~\ref*{fig:spinwave}D). Although the non-Hermitian blockade is weakened for $N=5,10,\dots$ due to near-resonant coupling channels, it still strongly suppresses a large fraction of excitations.

\end{document}